\documentclass[prd,twocolumn,nofootinbib,showpacs,superscriptaddress]{revtex4}
\usepackage{amsfonts}
\usepackage{amsmath}
\usepackage{amssymb}
\usepackage{upgreek}
\usepackage{bm}
\usepackage{dcolumn}
\usepackage{epsfig}
\usepackage{graphicx}
\usepackage{graphics}
\usepackage[latin1]{inputenc}
\usepackage{latexsym}
\usepackage{rotating}
\usepackage{hyperref}

\newcommand\be{\begin{equation}}
\newcommand\ba{\begin{eqnarray}}
\newcommand\ee{\end{equation}}
\newcommand\ea{\end{eqnarray}}

\newcommand{\mb}[1]{\mbox{\boldmath $#1$}}

\newcommand{\nn}{\nonumber}

\newcommand{\GW}{{\mbox{\tiny GW}}}
\newcommand{\GR}{{\mbox{\tiny GR}}}

\newcommand{\TT}{{\mbox{\tiny TT}}}
\newcommand{\MAT}{{\mbox{\tiny mat}}}
\newcommand{\STF}{{\mbox{\tiny STF}}}

\newcommand{\CS}{{\mbox{\tiny CS}}}

\newcommand{\MIN}{{\mbox{\tiny min}}}
\newcommand{\INC}{{\mbox{\tiny inc}}}
\newcommand{\APO}{{\mbox{\tiny apo}}}
\newcommand{\PERI}{{\mbox{\tiny peri}}}
\newcommand{\pont}{{\,^\ast\!}R\,R}
\newcommand{\met}{\mbox{g}}

\newcommand{\oone}{{}^{\mbox{\tiny $(1)$}}}
\newcommand{\otwo}{{}^{\mbox{\tiny $(2)$}}}
\newcommand{\lgw}{\lambda^{}_{\GW}}
\newcommand{\hgw}{h^{}_{\GW}}

\begin{document}
\title{Extreme- and Intermediate-Mass Ratio Inspirals\\
in Dynamical Chern-Simons Modified Gravity}

\author{Carlos F.~Sopuerta}
\affiliation{Institut de Ci\`encies de l'Espai (CSIC-IEEC), 
Facultat de Ci\`encies, Campus UAB, Torre C5 parells, 
Bellaterra, 08193 Barcelona, Spain.}

\author{Nicol\'as Yunes}
\affiliation{Princeton University, Physics Department, Princeton, NJ 08544, USA.}

\date{\today}

%%%%%%%%%%%%%%%%%%%%%%%%%%%%%%%%%%%%%%%%%%%%%%%%%
\begin{abstract} 

Chern-Simons (CS) modified gravity is a four-dimensional, effective theory that descends both from string theory 
and loop quantum gravity, and that corrects the Einstein-Hilbert action by adding the product of a scalar field 
and the parity-violating, Pontryagin density. 
The Chern-Simons modification deforms the gravitational field of spinning black holes, which is
now described by a modified Kerr geometry 
whose multipole moments deviate from the Kerr ones only at the fourth multipole, $\ell = 4$.
This paper investigates possible signatures of this theory in the gravitational wave emission produced
in the inspiral of stellar compact objects into massive black holes, both for intermediate- and
extreme-mass ratios.  
We use the {\emph{semi-relativistic}} approximation, where the trajectory of the small compact object 
is modeled via geodesics of the massive black hole geometry, while the gravitational waveforms are 
obtained from a multipolar decomposition of the radiative field.
The main Chern-Simons corrections to the waveforms arise from modifications to the geodesic trajectories, 
which in turn are due to changes to the massive black hole geometry, and manifest themselves as
an accumulating dephasing relative to the general relativistic case.
We also explore the propagation and the stress-energy
tensor of gravitational waves in this theory, using the short-wavelength approximation.  We find that, although this tensor
has the same form as in General Relativity, the energy and angular momentum balance laws are indeed modified 
through the stress-energy tensor of the Chern-Simons scalar field.  
These balance laws could be used to describe the inspiral through adiabatic changes in the orbital parameters, 
which in turn would enhance the dephasing effect.  
Gravitational-wave observations of intermediate- or extreme-mass ratio inspirals with advanced 
ground detectors or with the Laser Interferometer Space Antenna (LISA) could use such dephasing to test 
the dynamical theory to unprecedented levels, thus beginning the era of gravitational wave tests of 
effective quantum gravity theories.

\end{abstract}


\maketitle

%%%%%%%%%%%%%%%%%%%%%%%%%%%%%%%%%%%%%%%%%%%%%%%%%%%%%%%%%%%%%%%%%%%%%%%%%%%%%%
\section{Introduction}
\label{intro}

The experimental verification of symmetry breaking is one of the most powerful tools to understand in which direction
to extend the current canon toward more fundamental physical theories. For example, the experimental confirmation of 
the violation of charge conjugation, parity transformation and time-reversal symmetries in elementary particle 
interactions forced the improvement of the quantum field theory of particles into what is today the standard model. 
Similarly, violation of symmetries in gravitational interactions can push toward generalizations of General 
Relativity (GR) by providing the first experimental evidence of high-energy extensions.    

Gravitational parity violation can be tuned by the inclusion of a Pontryagin term in the Einstein-Hilbert 
action, which defines an effective, four-dimensional gravitational theory: Chern-Simons (CS) modified 
gravity~\cite{jackiw:2003:cmo}. In fact, the inclusion of such a term in the action is inescapable in 
four-dimensional compactifications of perturbative string theory ({\emph{i.e}}.~Type I, IIb, Heterotic, etc.) 
due to the Green-Schwarz anomaly-canceling mechanism~\cite{Polchinski:1998rr}. This fact can also be extended to 
the non-perturbative sector in the presence of Ramond-Ramond scalars (D-instanton charges) due to duality 
symmetries~\cite{Alexander:2004xd}. Such a term also arises naturally in loop quantum gravity when the 
Barbero-Immirzi parameter is promoted to a scalar field coupled to the Nieh-Yan invariant~\cite{Taveras:2008yf,Mercuri:2009zt}.

The action of Dynamical CS Modified Gravity (DCSMG) consists of the Einstein-Hilbert term plus the product of a scalar field and the 
Pontryagin density (the contraction of the Riemann curvature tensor with its dual), plus the action for this scalar field
and/or other matter fields. The correction proportional to the Pontryagin density modifies the 
field equations for the metric components 
by adding two extra terms to the Einstein equations: the so-called {\em C-tensor} and an stress-energy tensor
for the scalar field. The C-tensor depends on derivatives 
of the CS scalar and the contraction of the Levi-Civita tensor with covariant derivatives of the Ricci tensor 
and the dual Riemann tensor.  In addition, the variation of the action with respect to the CS scalar field 
leads to an equation of motion for this field, which is sourced by the Pontryagin density.

The CS gravitational modification has been investigated mostly in the non-dynamical framework, 
in which the scalar field is non-dynamical (there is no kinetic term for it in the action), and hence 
it is assumed to be an \emph{a priori} 
prescribed spacetime function. Such studies include an analysis of exact solutions~\cite{jackiw:2003:cmo,Guarrera:2007tu},
approximate solutions~\cite{jackiw:2003:cmo,Alexander:2007zg,Alexander:2007vt,Alexander:2007kv,Yunes:2008bu,Konno:2007ze}, 
matter interactions~\cite{Cantcheff:2008qn,Alexander:2008wi}, 
cosmology~\cite{Alexander:2006mt,Alexander:2004wk,Alexander:2004xd,Alexander:2007qe}, and astrophysical 
tests~\cite{Smith:2007jm,Konno:2008np,Yunes:2008ua}.  Non-dynamical CS modified gravity has been shown 
to be theoretically problematic in relation to Schwarzschild black hole perturbation 
theory~\cite{Yunes:2007ss}, the existence of stationary and axisymmetric solutions~\cite{Grumiller:2007rv}, 
and the uniqueness of solutions of the theory~\cite{Yunes:2009hc}. 

The detailed study of the dynamical formulation of CS modified gravity has only recently begun.  
This paper is the second in a series that deals with the details of the dynamics of CS modified, spinning black holes.
In the first paper~\cite{Yunes:2009hc}, henceforth Paper I, an approximate solution was found for 
a spinning black hole, using the slow-rotation approximation and a small-coupling approximation. The first type of approximation
restricts attention to black holes with small angular momentum per unit mass, while the second one allows one to 
search for small CS deformations of known GR solutions. The new solution corresponds to a deformation of the 
Kerr metric whose deviations fall off with a high power of the distance 
to the black hole.

In this paper, we concentrate on the study of intermediate and extreme-mass ratio inspirals (IMRIs and EMRIs respectively) 
in the context of DCSMG.  Such systems consist of a small compact object (SCO) 
(with masses in the range $1-30\,M^{}_{\odot}$) orbiting around a (spinning) massive black hole (MBH),
(with masses in the range $10^4-10^7 M_{\odot}$) in the case of EMRIs, and an intermediate-mass black
hole (IMBH) (with masses in the range $10^{2}-10^{4}\,M^{}_{\odot}$; see~\cite{Miller:2003sc} for a review on the evidence 
of the existence of IMBHs) in the case of IMRIs.  Another IMRI possibility would be that of an IMBH falling into a MBH, 
a system with obviously also an intermediate mass ratio (see~\cite{Miller:2008fi} and references therein). 
The mass ratios involved are then in the range $(10^{-2}-10^{-4})$ for IMRIs  and $(10^{-4}-10^{-7})$ for EMRIs.  

EMRIs (and IMRIs involving a IMBH-MBH binary) are important sources of gravitational 
waves (GWs) for future space detectors~\cite{Gair:2004ea} as the Laser 
Interferometer Space Antenna (LISA)~\cite{Vitale:2002qv,Danzmann:2003tv,Danzmann:2003ad,Prince:2003aa,lisa},
whereas IMRIs are important sources for second generation ground based detectors 
(see~\cite{Brown:2006pj,Mandel:2007hi}), such as Advanced LIGO~\cite{ligo} 
and Advanced VIRGO\cite{virgo}, and for future planned third generation detectors as the Einstein
Telescope~\cite{et}.  Both IMRIs and EMRIs are high-precision sources that can produce a high 
number of detectable GW cycles. Hence, GW observations of these systems 
will have a strong impact on astrophysics, cosmology, and
fundamental physics~\cite{Schutz:2009tz,Prince:2009uc,Miller:2009wv}, the latter of which we shall be concerned with here
(see~\cite{AmaroSeoane:2007aw} for a detailed account of  IMRI and EMRI  astrophysics and related science).

The construction of IMRI/EMRI waveforms for data analysis purposes is a difficult problem since the
accuracy needed for detection and extraction of physical parameters is quite high.  For the case of EMRIs,
techniques for constructing sufficiently accurate templates for {\emph{detection}} is currently underway, mainly through 
the use of the adiabatic 
approximation~\cite{Apostolatos:1993nu,Cutler:1994pb,Kennefick:1995za,Kennefick:1998ab,Mino:2003yg,Hughes:2005qb,Drasco:2005kz,Sago:2005gd,Ganz:2007rf,Fujita:2009rr}. 
Methods to build sufficiently accurate templates for {\emph{parameter estimation}} has not yet been fully 
developed, the main difficulty being that one requires a more precise treatment of the self-gravity 
of the SCO and its impact on the gravitational waveform.  For IMRIs, one might have to incorporate 
the finite-size effects of the SCO on the waveforms, since the mass ratio is not so extreme.

In this article, we will use a simpler method to describe EMRIs, the so-called {\emph{semi-relativistic}} 
approximation~\cite{Ruffini:1981rs}, in which the motion of the SCO is taken to be geodesic on
the MBH spacetime (Kerr in GR and modified Kerr in DCSMG) 
and the GWs are computed using a multipolar decomposition~\cite{Thorne:1980rm}.
In the radiation zone, the different multipoles are fully determined by the trajectory of the SCO and its derivatives.
Such a scheme therefore neglects radiation-reaction effects that scale with the square of the mass ratio of the system. 
One can add these effects by using different types of approximations that provide expressions for the change of 
the {\em constants} of motion in terms of the properties of the GWs 
emitted~\cite{Barack:2003fp,Gair:2005is,Gair:2005hu,Babak:2006uv}. Although the semi-relativistic approximation is probably not 
sufficient to create a template bank for parameter estimation, it will suffice here to understand the type of 
corrections induced in CS modified EMRIs and IMRIs.

We begin by considering the trajectory of the SCO in the MBH background.
We find that the motion of a test particle in DCSMG is 
{\emph{exactly}} described by the geodesic equations, as in GR. 
Trajectory modifications are thus produced entirely by the modified geometry of the MBH background, 
which is here described by the modified Kerr solution found in Paper I. These modifications include
corrections to the geodesic equations and to the fundamental frequencies of motion.  
Since the CS correction to the MBH background is a high-order, strong-field effect,  
the multipoles of the hole are not corrected up to the $\ell=4$ multipole, suggesting that a test of the Kerr geometry 
might be difficult with LISA.   
This theory is thus an interesting illustration of how the geometry of a MBH can be changed by physically 
well-motivated curvature corrections to the Einstein-Hilbert action.

We continue with a discussion of GW generation in DCSMG, which for EMRIs and IMRIs can be studied using black
hole perturbation theory.  We see that the perturbations, and hence the GW emission, are indeed corrected by 
the C-tensor and the stress-energy tensor of the CS scalar field. Expanding in the mass ratio and in the CS coupling 
constants that act as CS deformation parameters, we find that the second-order corrections to GW emission are
the usual GR radiation-reaction effect (proportional to the square of the mass ratio) and a new CS correction  
(proportional to the product of the mass-ratio and the CS coupling constants).  Depending on the 
strength of these constants, this modification could be much smaller, comparable, or bigger
than radiation-reaction effects.  
In this paper, as a first exploratory study of these questions, we shall neglect these second-order effects, and thus,  
we shall model GW emission through the usual GR multipolar expansion. Modifications to 
GW emission, hence, originate from corrections to the geodesic trajectories of the SCO, 
which in turn arise due to the modified MBH geometry.

\begin{figure}[htp]
\begin{center}
\includegraphics[scale=0.325,clip=true]{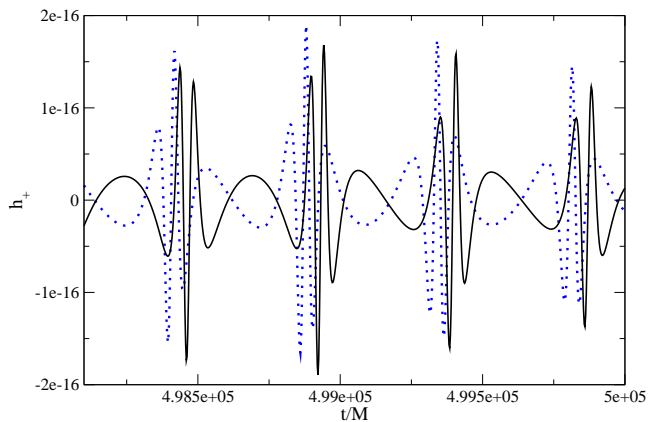}
\caption{\label{dephaseHpoft-zoom} Plus-polarized waveform as a function of time using a Kerr background 
(dotted blue) and the CS deformed metric (solid black). The pericenter is at $r = 6 M$ (with $M = 4.5 \times 10^6 
M^{}_{\odot}$) the inclination angle is $0.3$ radians, the spin 
angular momentum is $J/M^{2} = 0.4$, while the CS coupling strength is set at $\xi/M^{4} = 0.4$. 
For more details on this system see Sec.~\ref{test-sys} and for further details on the precise definition of 
the CS coupling strength see Sec.~\ref{CS-basics}.}
\end{center}
\end{figure}

With these tools at hand, we then proceed to solve the geodesic equations for a set of EMRI systems and compare 
the GWs generated in GR with those generated in DCSMG.  
Figure~\ref{dephaseHpoft-zoom} shows the changes in the waveforms, essentially an accumulating dephasing, for 
the last eleven minutes of the plus-polarized waveform after $128$ days of evolution for the system described 
in the caption. The GR waveforms are here denoted with a dotted blue line, while the DCSMG waveforms are 
denoted with a black solid line. 
The amount of CS dephasing depends on the specific type of orbit and the magnitude of the CS correction, where 
we generically find a strong effect on the waveform for orbits that spend the longest close to the MBH. 

%%%%%%%%%%
%%%%%%%%%%

Our study suggests that a GW observation of highly relativistic IMRIs and EMRIs 
could place constraints on the dynamical theory that are orders of magnitude larger than current binary pulsar 
ones~\cite{Yunes:2009hc}.  This constraint depends strongly on how relativistic the system is, as well as on the 
signal-to-noise ratio of the event, and on the total mass of the system.  Including radiation-reaction effects 
should allow us to search, for the first time, for radiative effects associated with the CS scalar field in the 
neighborhood of compact objects, which is simply not possible with binary pulsars (see~\cite{lrr-2006-3,lrr-2008-9} 
for detailed information on tests of gravitational theories, in particular for tests with pulsars).  To confirm 
these expectations a detailed analysis using appropriate data analysis tools is required, but shall not be
performed here.

Remarkably, a physically well-motivated curvature-squared correction to GR, like the one studied here, 
seems to lead to spinning black hole solutions and GWs that resemble 
the GR prediction quite closely. This fact suggests that GW detectors should be able to detect GWs, irrespective 
of whether GR or CS waveform templates are employed.  As long as the precise nature of the massive compact object 
or the structure of the true gravitational theory does not modify the GR predictions enough, as we show is the case in 
DCSMG, the current data analysis algorithms should be able to extract the signals by 
using purely GR templates and performing incoherent searches of the data in short segments (e.g.~roughly three 
weeks).  However, in our case, the CS modification may prevent us from {\emph{connecting}} these segments together and 
associate them to a specific EMRI or IMRI system.  

Although the study of GWs emitted by EMRIs associated to non-Kerr backgrounds is not 
new~\cite{Collins:2004ex,Glampedakis:2005cf}, the analysis presented here is the first one to consider an alternative to
the Kerr metric that derives from a concrete and physically well-motivated alternative theory of gravity (based on
high-order curvature corrections to GR).  In contrast to arbitrary quasi-Kerr deformations, the black hole solution 
considered here deviates from Kerr significantly {\emph{only in the strong field}} region and only through gravitomagnetic 
components.  Thus, mismatches between GW signals of EMRIs in this background (and in DCSMG) and GWs in Kerr (and in GR)
are dominated exclusively by truly strong-field effects and not by subleading, weak-field corrections.  
Tests of such modified theories seem to require not only the detection of a sufficiently large number of segments, 
but also the reconstruction of a unique GW signal.

We conclude this paper with a discussion of the role of radiation-reaction effects in DCSMG through the 
short-wavelength approximation.  In particular, we investigate how one could introduce these effects in the 
adiabatic approximation (that is, when the rate of change of the orbital parameters due to GW emission is much 
smaller that the typical orbital periods) in order to construct truly inspiraling IMRI/EMRI waveforms in DCSMG.   
We find that the effective GW stress-energy tensor, the Isaacson tensor, and thus the GW stress-energy loss, 
is the {\emph{same}} in DCSMG as in GR.   The backreaction of this loss on the background, however, is different 
in DCSMG because GW losses must here be supplemented by the stress-energy distribution of the CS scalar field. 
Truly inspiraling IMRI/EMRI waveforms can then be constructed in DCSMG by supplementing the semi-relativistic approximation
with adiabatic changes of the orbital parameters due to {\emph{both}} GW and CS scalar field stress-energy losses. 

The details of this study are organized as follows: Sec.~\ref{CS-basics} discusses the basics of DCSMG, its linearized 
version, the study of GW polarization in this theory, and reviews the small-coupling approximation; 
Sec.~\ref{motion} describes test-particle motion in DCSMG, including the expression of the modified Kerr solution,
and the derivation of the equations for timelike geodesics.  We also discuss how the fundamental frequencies of geodesic
motion are modified due to the changes in the MBH metric; Sec.~\ref{h-quad} describes GW generation in the modified theory and
the type of semi-relativistic approximation that we use; 
in Sec.~\ref{semi-class} we apply the previous results to construct numerically the trajectories and waveforms;
Sec.~\ref{radiationreaction} discusses GW propagation in the short-wavelength approximation and shows that the GW energy momentum tensor 
and emission in DCSMG is the same as in GR.   We also calculate the stress-energy emission due to the CS scalar 
field and the modification to the background (averaged) scalar field due to terms quadratic in the perturbation of 
the Riemann tensor; Sec.~\ref{conclusions} concludes and discusses future research directions.

We use the following conventions throughout this work. Greek letters and a semicolon are used to denote indices and covariant
differentiation respectively on the $4$-dimensional spacetime.   We denote covariant differentiation with respect 
to the background metric by $\bar\nabla^{}_\mu B^{}_{\nu}$ or by $B^{}_{\nu|\mu}$.  Partial differentiation with respect to the
coordinate $x^{\mu}$ is denoted as $\partial^{}_{\mu} B^{}_{\nu}$ or $B^{}_{\nu,\mu}$.
Symmetrization and antisymmetrization is denoted with parenthesis and square brackets around the indices respectively, 
such as $A^{}_{(\mu\nu)}:= [A^{}_{\mu\nu} + A^{}_{\nu\mu}]/2$ and $A^{}_{[\mu\nu]} := [A^{}_{\mu\nu} - A^{}_{\nu\mu}]/2$. 
We use the metric signature $(-,+,+,+)$ and geometric units in which $G = c = 1$.

%%%%%%%%%%%%%%%%%%%%%%%%%%%%%%%%%%%%%%
\section{Chern-Simons Modified Gravity}
\label{CS-basics}

In this section we describe the formulation of CS modified gravity that we shall
employ and establish some basic notation (see also Paper~I).  We begin with a discussion of the 
basics of the modified theory, but refer the reader to Paper~I or the upcoming review~\cite{review} for further details.
We continue with a review of the linearized theory in DCSMG.  Then we study basic properties of GW polarization in DCSMG,
illustrated with plane waves.  Finally we introduce the small-coupling approximation.

%----------------------------------------------------------------------------------------------------
\subsection{General Formulation}
The starting point is the action of DCSMG:
\ba
\label{full-action}
S &=& S^{}_{\rm EH} + S^{}_{\rm CS} +  S^{}_{\vartheta} + S^{}_{\rm mat},
\ea
where
\ba
\label{actions}
S^{}_{\rm{EH}} &=& \kappa \int d^4x  \; \sqrt{-\met}  \; R\,, 
\nonumber \\
S^{}_{\rm{CS}} &=& \frac{\alpha}{4} \int d^4x \; \sqrt{-\met} \quad
\vartheta \; \pont\,,
\nonumber \\
S^{}_{\vartheta} &=& - \frac{\beta}{2} \int d^{4}x \; \sqrt{-\met} \; \left[ \met^{\mu \nu}
\left(\nabla_{\mu} \vartheta\right) \left(\nabla_{\nu} \vartheta\right) + 2 V(\vartheta) \right]\,,
\nonumber \\
S^{}_{\textrm{mat}} &=& \int d^{4}x \; \sqrt{-\met} \; {\cal{L}}_{\textrm{mat}}\,,
\ea
where $\kappa = (16 \pi G)^{-1}$ is the gravitational constant; $\alpha$ is the coupling constant of the
CS scalar field $\vartheta$ with the parity-violating Pontryagin density, $\pont$, given by
\be
\pont  := R^{}_{\alpha \beta \gamma \delta}
{\,^\ast\!}R^{\alpha \beta \gamma \delta} =
\frac{1}{2} \epsilon^{\alpha \beta \mu \nu} 
R^{}_{\alpha \beta \gamma \delta}
R^{\gamma \delta}{}^{}_{\mu \nu}\,,
\label{CSmodification}
\ee
where the asterisk denotes the dual tensor, constructed using the antisymmetric Levi-Civita tensor 
$\epsilon^{\alpha\beta\mu\nu}$; $\beta$ is a constant that determines the gravitational strength of 
the CS scalar field stress-energy distribution.  The different terms in Eq.~\eqref{full-action} correspond 
to the following: the first one is the Einstein-Hilbert action; the second one is the CS gravitational correction; 
the third one is the CS scalar field action, which contains a kinetic and a potential term $V(\vartheta)$, both 
of which distinguishes the dynamical formulation from previous ones; 
and the fourth one is the action corresponding to matter degrees of freedom.  
 
Upon variation of the action with respect to the metric and the CS scalar, we obtain the field equations
of DCSMG:
\ba
G^{}_{\mu \nu} + \frac{\alpha}{\kappa} C^{}_{\mu \nu}  &=& \frac{1}{2 \kappa} \left(T_{\mu \nu}^{\MAT} 
+ T_{\mu \nu}^{(\vartheta)}\right)\,,  \label{EEs} \\
\beta \square \vartheta &=& \beta \frac{d V}{d\vartheta} - \frac{\alpha}{4} \pont\,,
\label{EOM}
\ea
where $T_{\mu \nu}^{\MAT}$ is the matter stress-energy tensor and $T_{\mu \nu}^{(\vartheta)}$ is the stress-energy of the 
CS scalar field, given by 
\be
T_{\mu \nu}^{(\vartheta)} = \beta \left[ (\nabla_{\mu} \vartheta) (\nabla_{\nu}\vartheta) - \frac{1}{2} 
\met_{\mu \nu}(\nabla^{\sigma} \vartheta) (\nabla_{\sigma} \vartheta) -  \met_{\mu \nu}V(\vartheta) \right]\,.
\label{vartheta-Tab}
\ee
In this work we will study only the case 
in which the potential term associated with the CS scalar field is assumed to vanish\footnote{In 
string theory, the $\vartheta$ field corresponds to the axion, which possess a shift symmetry and a vanishing potential 
along certain {\em flat} directions. The axion is a moduli field, which, when stabilized by supersymmetry 
breaking~\cite{Gukov:2003cy}, acquires a non-flat potential. Such potential would arise at the scale of supersymmetry 
breaking, which is much larger than the gravitational scale.} ($V=0$).
The tensor $C^{\mu \nu}$ (the so-called {\em C-tensor}) in Eq.~\eqref{EEs} can be 
split into two parts, $C^{\mu \nu} = C^{\mu \nu}_{1} + C^{\mu \nu}_{2}$, where
\ba
\label{Ctensor}
C^{\alpha \beta}_{1} &=& \left(\nabla^{}_{\sigma} \vartheta\right) \epsilon^{\sigma \delta \nu(\alpha}
\nabla^{}_{\nu}R^{\beta)}{}^{}_{\delta}\,, \nonumber \\
C^{\alpha \beta}_{2} &=& \left(\nabla^{}_{\sigma}\nabla^{}_{\delta}\vartheta \right) 
{\,^\ast\!}R^{\delta (\alpha \beta)\sigma}\,.
\ea

As we have mentioned, there are two conceptually distinct CS modified gravity theories: a dynamical version and a 
non-dynamical version. In the former, the quantities $\alpha$ and $\beta$ are arbitrary, and the field equations are 
the ones we have given in Eqs.~\eqref{EEs} and~\eqref{EOM}.  The non-dynamical version is characterized by the choice
$\beta = 0$ at the level of the action, and thus the evolution equation for the CS scalar becomes a differential 
constraint on the space of allowed solutions, the so-called Pontryagin constraint $\pont = 0$.

The non-dynamical theory presents a certain number of difficulties that make it less physically interesting than the 
dynamical formulation~\cite{Yunes:2007ss,Grumiller:2007rv,Yunes:2009hc}. In the former, one can show that single-parity 
perturbation modes lead to an overconstrained system of perturbed equations, disallowing generic quasinormal 
ringing~\cite{Yunes:2007ss}. Moreover, spinning black hole solutions were found to be strongly restricted by the 
Pontryagin constraint, in some cases also leading to an overconstrained system~\cite{Grumiller:2007rv}. 
Finally, the freedom in the choice of the CS scalar field was shown to lead to non-unique solutions with infinite 
energy scalars in Paper I~\cite{Yunes:2009hc}.       

Most of the difficulties and arbitrariness of the non-dynamical theory can be avoided by adopting the dynamical framework, 
i.e.~DCSMG theory, where the freedom regarding the CS scalar field reduces to the choice of initial conditions. 
The stationary state of the CS scalar has been seen to be independent of the initial conditions, being determined 
only by the metric through the source term in the evolution equation for $\vartheta$~\cite{Yunes:2009hc}. 
For these reasons, we choose here to study the dynamical formulation.

%----------------------------------------------------------------------------------------------------
\subsection{The Linearized Theory}
\label{lin-th}

In perturbation theory we can split the spacetime metric as the sum of a background metric 
$\bar{\met}^{}_{\mu \nu}$ (we use the overbar to denote objects associated with the background) 
and a metric perturbation $h_{\mu \nu}$:
\be
\met^{}_{\mu\nu} = \bar{\met}^{}_{\mu\nu} + \epsilon \;  h^{}_{\mu\nu}\,, \label{splitting}
\ee
where $\epsilon$ is a book-keeping perturbative parameter, associated to the smallness of the metric 
perturbation relative to the background, which we use to label the order of the approximation.
In this section and in other parts of this paper, we simplify equations by using the Lorenz gauge
\be
\left(h^{\mu\nu} - \frac{1}{2}h \bar\met^{\mu\nu}\right)^{}_{|\nu}  = 0\,, \label{lorenzgauge}
\ee
where $h = \bar{\met}^{\mu \nu} h_{\mu \nu}$ is the trace of the metric perturbation with respect to the background.
Since we shall be at first interested in computing the DCSMG corrections to  
gravitational radiation measured by observers in the radiation zone, we 
shall restrict ourselves to a flat background, $\bar{\met}^{}_{\mu \nu} =  \eta_{\mu \nu}$, which is a good approximation sufficiently far away
from the source. In the radiation zone, we can further restrict the gauge
by supplementing it with the trace-free condition
\be
h = 0\,, \label{tracelesscondition}
\ee
leading to a transverse-traceless gauge far away from the source.

In this gauge, the linearized Riemann, dual Riemann, and Ricci tensors are ~\cite{Carrol}: 
\ba
R_{\mu \nu \rho \sigma} &=& h_{\sigma [ \mu,\nu] \rho} - h_{\rho [\mu, \nu] \sigma}\,, \\
{\,^\ast\!}R^{\delta \alpha \beta \gamma} &=& \bar\epsilon^{\,\delta \mu \nu \alpha} h_{\nu}{}^{[\beta,\gamma]}{}_{\mu}\,, \\
R_{\mu \nu} & = & - \frac{1}{2} \square_{\eta}\, h_{\mu \nu}\,,
\ea
respectively, where $\square_{\eta} = \eta^{\mu \nu} \partial_{\mu} \partial_{\nu}$ is the flat-space 
D'Alembertian and $\bar{\epsilon}^{\,\mu \nu \alpha \beta}$ is the flat-space Levi-Civita symbol. 
The linearized C-tensor in this gauge is then 
\ba
C_{\mu \nu} &=& - \frac{1}{2} \;  (\partial_{\sigma} \vartheta)  \; 
  \bar\epsilon^{\,\sigma \delta \xi}{}_{(\mu} \; \square_{\eta} h_{\nu) \delta,\xi} \,,
\nonumber \\
&-& \frac{1}{2} (\partial_{\sigma}{}^{\gamma} \vartheta)  \;
\bar \epsilon^{\,\sigma \delta \xi}{}_{(\mu|} 
\left[ h_{\xi \gamma,|\nu) \delta} - h_{\nu) \xi,\gamma \delta} \right] \,.
\ea
Combining these linearized expressions, we find that the linearized modified field equations 
in trace-reversed form become
\ba
 \label{lin-mod-FE}
- \frac{1}{\kappa} \bar{T}_{\mu \nu}^{\vartheta} &=&
\square_{\eta} h_{\mu \nu} 
+ \frac{\alpha}{\kappa} (\partial_{\sigma} \vartheta) \bar\epsilon^{\,\sigma \delta \xi}{}_{(\mu} 
\square_{\eta} h_{\nu) \delta,\xi}  \nonumber \\
&+& \frac{\alpha}{\kappa} (\partial_{\sigma}{}^{\gamma} \vartheta) \bar\epsilon^{\,\sigma\delta \xi}{}_{(\mu} 
\left[ h_{\xi \gamma,|\nu) \delta} - h_{\nu) \xi,\gamma \delta} \right]\,, \\
 \beta \square \vartheta &=& -\frac{\alpha}{2} \bar{\epsilon}^{\,\alpha \beta \mu \nu} h_{\alpha \delta,\gamma \beta} 
h_{\nu}{}^{[\gamma,\delta]}{}_{\mu}\,, \label{CS-FE}
\ea
where $\bar{T}_{\mu \nu}^{\vartheta} = T_{\mu \nu}^{\vartheta} - (1/2) \eta^{}_{\mu \nu} T^{\vartheta}$ is the 
trace-reversed stress-energy tensor of the  CS scalar-field.  We see clearly that the metric perturbation 
is sourced by $\bar{T}_{\mu \nu}$, which implies that the contribution of this source must also be of 
order ${\cal{O}}(\epsilon)$, while the scalar field is sourced by the Pontryagin density that is of 
${\cal{O}}(\epsilon^{2})$ in the radiation zone.   Note in this regard, that we have not yet introduced
any perturbative expansion for the CS scalar field.

%----------------------------------------------------------------------------------------------------
\subsection{Gravitational-Wave Polarizations}\label{gwpolarization}
Since we are dealing with a different theory of gravity, one must be specially 
careful of not assuming GW properties that hold only in GR.  In this sense, a question 
of particular relevance that arises with alternative theories of gravity refers to
the number of independent GW polarizations. DCSMG possess an extra degree of freedom
described by the CS scalar field, and thus, additional GW polarizations
could in principle be present.  This issue can be investigated by following the pioneering work of~\cite{Eardleyprd,Eardleyprl}, where 
a formalism is presented to study far-field gravitational radiation in any metric theories of gravity.
One then finds that only six possible independent GW polarizations are possible, leading to
a suitable classification (the so-called $E(2)$ classification) of alternative theories.  

In this framework one focuses on the propagation of monochromatic plane waves in the 
radiation zone (i.e. assuming weak-fields) and their effect on test masses.  This can be
done by considering the geodesic deviation equation (see e.g.~\cite{Misner:1973cw,Wald:1984cw}), which
describes the relative acceleration between nearby geodesics: 
\be
U^{\rho}\nabla^{}_{\rho}(U^{\sigma}\nabla^{}_{\sigma}X^{\mu}) := A^{\mu} = 
R^{\mu}{}_{\nu\rho\sigma} U^{\nu}U^{\rho} X^{\sigma}\,, \label{deviationequation}
\ee
where $U^{\mu}$ is the test mass four-velocity and $X^{\mu}$ is the so-called deviation vector, which
describes the displacement between test masses. Different GW polarizations will induce a different 
deviation effect on test masses, that can be classified via the structure of the Riemann tensor.
Thus, the $E(2)$ classification sorts metric theories based on the vanishing or non-vanishing of 
the Newman-Penrose scalars (in an appropriate Newman-Penrose null basis), 
associated with the Riemann curvature tensor, which can be shown to describe GW polarizations.  

Applying this procedure to DCSMG (one can use the formulae of the linear theory presented above),
we find that only the two GW polarizations also present in GR are observable 
far away from sources.  Such a study had not
yet been performed in DCSMG, although several investigations exist in the non-dynamical theory: 
~\cite{jackiw:2003:cmo} proved the above result in the non-dynamical theory for a linearly time-dependent
scalar field, while~\cite{Alexander:2007kv,Yunes:2008bu} extended this result to generic time-dependent scalars. 

To illustrate this fact in the non-linear regime, let us consider the
propagation of plane-fronted gravitational waves ({\em pp-waves}) in DCSMG 
(see~\cite{Grumiller:2007rv} for pp-waves in the non-dynamical formalism).
The line element for pp-waves can be written as (see e.g.~\cite{Stephani:2003tm}):
\be
ds^{2} = -2dudv - H(u,x,y)du^{2} + dx^{2} + dy^{2}\,, \label{ppwaves}
\ee
where $u=t-z$ and $v=t+z$ are retarded and advance null coordinates respectively (the waves propagate
along the $z$ axis), the wave fronts
are given by the null planes $u={\textrm{const.}}$, and all physical information is encoded in the scalar $H$. Indeed, 
using a Newman-Penrose null basis~\cite{Newman:1961qr,Stephani:2003tm} 
adapted to pp-waves ($\mb{k}=\partial^{}_{v}\,$, $\mb{\ell}=\partial^{}_{u}-(H/2)\partial^{}_{v}$, $\mb{m}=(\partial^{}_{x}
+i\,\partial^{}_{y})/\sqrt{2}$) one finds that the only two non-vanishing Newman-Penrose complex curvature scalars are:
\be
\Psi^{}_{4} = \frac{1}{4}(\partial^{2}_{x}-\partial^{2}_{y} + 2\,i\,\partial^{2}_{xy})H\,,~~
\Phi^{}_{22} = \frac{1}{4}(\partial^{2}_{x}+\partial^{2}_{y})H\,. \label{npcurvaturescalars}
\ee
Considering an arbitrary CS scalar field, $\vartheta = \vartheta(u,v,x,y)$ and the
metric in Eq.~\eqref{ppwaves}, the modified field equations [Eqs.~\eqref{EEs} and~\eqref{EOM}]
force us to
\be
\vartheta=\vartheta(u)\,,\qquad
(\partial^{2}_{x}+\partial^{2}_{y})H = \frac{\beta}{\kappa}\dot\vartheta^{2}\,, \label{ppdcsmg}
\ee
where $\vartheta(u)$ is an arbitrary function and $\dot\vartheta = d\vartheta/du$.  From Eqs.~\eqref{npcurvaturescalars} 
and~\eqref{ppdcsmg} we see that 
in GR ($\beta=0$) the only non-vanishing Weyl complex scalar is $\Psi^{}_{4}$, which encodes the two GR GW polarizations.  
In DCSMG, there is an extra degree of freedom due to the real Ricci scalar $\Phi^{}_{22}$, which describes a transverse GW whose
effect on a ring of test particles is to either expand or contract it maintaining a circular shape (a so-called {\em breathing} mode)~\cite{Eardleyprd}.   
The non-vanishing of $\Psi_{4}$ and $\Phi_{22}$ suggests naively that DCSMG is of type $N^{}_{3}$ (like the Dicke-Brans-Jordan 
theory~\cite{Brans:1961sx}), but on closer inspection $\Phi_{22}$ is 
of ${\cal{O}}(\dot\theta^{2})$, and thus the theory reduces to type $N^{}_{2}$ (the same as GR) 
in the weak-amplitude approximation that is required in the $E(2)$ classification. 

This new polarization state can be studied more cleanly by considering {\em plane} waves~\cite{Baldwin:1926pw}, 
a special class of pp-waves characterized
by a scalar $H$ that is quadratic in $x$ and $y$ ($\Psi^{}_{4}$ and $\Phi^{}_{22}$ are independent of $x$ and $y$
and the metric has a five-dimensional Lie group of Killing symmetries~\cite{Stephani:2003tm}). 
Terms linear in $x$ and $y$ can be removed by coordinate 
transformations that leave invariant the line element in Eq.~\eqref{ppwaves}. The solution to Eq.~\eqref{ppdcsmg} 
can then be written as
\be
H(u,x,y) = A(u)(x^2-y^2) + 2B(u)xy + C(u)(x^2+y^2)\,,
\ee
where $A(u)$ and $B(u)$ are arbitrary functions and $C(u) = (\beta/(4\kappa))\dot\vartheta^2$. 
From~\eqref{npcurvaturescalars} one finds that $\Psi^{}_{4} = A+i\,B$ and $\Phi^{}_{22} = C$, 
clearly showing that the extra non-GR polarization corresponds to a breathing mode. 
In the weak-amplitude regime, however, $\vartheta^2 \simeq 0$, 
and the effect of this extra polarization is negligible, thus establishing the result that DCSMG only
dominantly excites the same two GW polarization states excited in GR.

%----------------------------------------------------------------------------------------------------
\subsection{The Small-Coupling Approximation}\label{smallcouplingapprox}

This approximation consists of expanding the modified field equations 
in the dimensionless parameter $\zeta = {\cal{O}}(\xi/M^{4})$, where $\xi := \alpha^{2}/(\beta \kappa)$ and $M$ is 
a characteristic mass associated with the particular system under consideration.   To that end, we assume that
$\zeta$ is a {\em small} perturbation parameter associated with the CS gravitational modifications and the
physical system under study.  As we did in the linearized theory, $\zeta$ will be also used as a book-keeping
parameter for labeling the different perturbative orders, but we will set it to unity at the end of our
calculations.   Combining this approximation with the linear one we can set up a two-parameter perturbative scheme,
in $\zeta$ and $\epsilon$,
that in the context of the DCSMG was introduced in Paper I and we refer to this work
for details (for a general discussion about multi-parameter perturbation theory 
see~\cite{Bender}, and in the context of GR 
see~\cite{Yunes:2005nn,Bruni:2002sm,Sopuerta:2003rg,Passamonti:2004je,Passamonti:2005cz}).   As in Paper I, 
the order counting shall assume that $\beta = {\cal{O}}(\alpha)$ such that $\alpha/\beta = {\cal{O}}(1)$ 
and $\xi = {\cal{O}}(\alpha/\kappa)$ without loss of generality.

The metric perturbations $h^{}_{\mu\nu}$ and the CS scalar field $\theta$ can then be expanded as
\ba
\label{h-exp}
h_{\mu \nu} &=& \sum^{}_{a,b} \epsilon^{a}\,\zeta^{b}\, h_{\mu \nu}^{(a,b)}\,,
\qquad
\vartheta = \sum^{}_{a,b} \epsilon^{a}\, \zeta^{b}\, \vartheta^{(a,b)}\,,
\ea
where $h^{(a,b)}_{\mu \nu}$ and $\vartheta^{(a,b)}$ stand for the perturbations of ${\cal{O}}(\xi^{a},\zeta^{b})$
and $a+b\geq 1$. Introducing these expansions in the field equations one can then set up a boot-strapping method 
that consists of first solving the evolution equation for the CS scalar field and then using this solution in the 
modified Einstein field equations to solve for the CS correction to the metric perturbations. 
In order to prevent the existence of metric perturbations whose dominant behavior is dramatically different 
from GR, we shall immediately set $h_{\mu \nu}^{(0,a)} = 0$ for all $a$, as we are here searching for CS-{\em deformations} 
of GR solutions.

%%%%%%%%%%%%%%%%%%%%%%%%%%%%%%%%%%%%%%
\section{Test Particle Motion}\label{motion}

The first approximation to the dynamics of IMRI/EMRIs is to treat the SCO as a point particle which, according
to GR follows geodesic motion. In this section we study how this picture is modified in DCSMG, from the
change in the geometry of the MBH to the change in the structure of the geodesic equations.

%----------------------------------------------------------------------------------------------------------------
\subsection{MBH Geometry}\label{backgroundquantities}

In GR, the geometry of a spinning MBH is described by the Kerr metric, but in the study of CS gravitational 
modifications it is well-known that this is no longer the case~\cite{Grumiller:2007rv}.
Recently, the corrections to the gravitational field of a spinning black hole have 
been found~\cite{Yunes:2009hc} in DCSMG using the slow-rotation 
approximation\footnote{After the publication of this work, 
other researchers~\cite{Konno:2009kg} have arrived at the same solution as that presented first in Paper I, 
suggesting a certain robustness and uniqueness of Eq.~\eqref{sol:metric_elements}.}, $a/M \ll 1$ 
(where $M$ refers to the MBH mass and $a$ is the MBH spin parameter), and the small-coupling 
approximation up to second order in $a/M$ and $\zeta$.  The form of the non-vanishing metric components using 
coordinates in which the GR part of the metric is Kerr in Boyer-Lindquist coordinates 
$(t,r,\theta,\phi)$ is:
\ba
\label{sol:metric_elements}
\bar{\met}^{}_{tt} &=& - \left(1 - \frac{2 M r}{\rho^{2}} \right) \,, 
\nonumber \\
\bar{\met}^{}_{t\phi} &=& - \frac{2 M a r}{\rho^{2}} \sin^{2}{\theta} 
\nonumber \\
&+& \frac{5}{8} \frac{\xi}{M^{4}} \frac{a}{M} \frac{M^{5}}{r^{4}} \left(1 + \frac{12 M}{7 r} 
+ \frac{27 M^{2}}{10 r^{2}} \right) \sin^{2}{\theta}\,,
\nonumber \\
\bar{\met}^{}_{rr} &=& \frac{\rho^{2}}{\Delta}\,,
\qquad
\bar{\met}^{}_{\theta \theta} = \rho^{2}\,,
\qquad
\bar{\met}^{}_{\phi \phi} = \frac{\Sigma}{\rho^{2}} \sin^{2}{\theta}\,,
\ea
where $\rho^{2} = r^{2} + a^{2} \cos^{2}{\theta}$, $\Delta = r^{2} f + a^{2}$, $f = 1 - 2 M/r$ 
and $\Sigma = (r^{2} + a^{2})^{2} - a^{2} \Delta \sin^{2}{\theta}$. The polar angle $\theta$ is not to be 
confused with the CS scalar field $\vartheta$. Technically, the expressions in
Eq.~\eqref{sol:metric_elements} are valid only up to second order in $a/M$ and $\zeta$, but we
have here chosen to resum the GR sector, by adding the appropriate high-order uncontrolled remainders.
In addition, the expression for the CS scalar field (using the same approximations) is:
\be
\bar{\vartheta} =  \frac{5}{8} \frac{\alpha}{\beta} \frac{a}{M} \frac{\cos(\theta)}{r^2} 
\left(1 + \frac{2 M}{r} + \frac{18 M^2}{5 r^2} \right)\,.
\label{back-theta}
\ee
The metric in Eq.~\eqref{sol:metric_elements} is stationary and axisymmetric, and the CS scalar
field has the same symmetries.  Notice that this CS scalar decays as $r^{-2}$ in the far-field 
and thus it possess a finite energy. 

One could in principle attempt to calculate the ${\cal{O}}[(a/M)^{3},\zeta]$ corrections to the metric. 
These corrections, however, are difficult to find because they involve modifications to all 
metric components, since the modified field equations must include the scalar stress-energy 
tensor to this order. We shall not consider these corrections here and work only to leading 
order with the metric of Eq.~\eqref{sol:metric_elements}. Needless to say, if the equivalent metric
for rapidly rotating MBHs were found, one could repeat the analysis of this paper for that 
background spacetime.

The non-{\em Kerrness} of Eq.~\eqref{sol:metric_elements} allows us to study DCSMG modifications 
to the multipolar structure of a spinning black hole, as this may illustrate
generic corrections of higher-order curvature extensions of GR. 
The multipolar structure of Kerr is fully determined by {\emph{only}} two multipole moments: 
the mass monopole and the current dipole.  All others are related to these via the simple
relation~\cite{hansen:46}: $M_{\ell} + i S_{\ell} = M \left(i a\right)^{\ell}$, where 
$\{M_{\ell}\}^{}_{\ell=0,\ldots,\infty}$ and $\{S_{\ell}\}^{}_{\ell=0,\ldots,\infty}$ are 
the mass and current multipole moments respectively.   
In DCSMG, the leading-order modification to this relation occurs for the $S^{}_{4}$ multipole, 
as one can see by employing the multipolar formalism of~\cite{Thorne:1980rm} (see also~\cite{Gursel}), 
and noting that the only metric  component that is CS modified scales as $r^{-4}$ for $M/r \ll 1$. 

DCSMG, therefore, preserves the idea of the no-hair (or two-hair) conjecture that spinning
black holes are determined by the mass and spin parameters, but it introduces a $4$-pole (or hexadecapole) correction. 
The reason why DCSMG does not violate the conjecture is because, apart from mass and spin, the other parameters
appearing in the MBH metric are introduced through the parameter $\xi$, which consists of fundamental
coupling constants of the theory that are fixed and non-tunable (i.e.~these constants are analogous to the Newtonian 
gravitational constant G).  Instead, the modifications introduced by DCSMG are such that the standard GR formulae to 
relate mass and angular momentum to all multipole moments of the solution does not hold.

The high $\ell$-number of the modification suggests that a GW test of the MBH spacetime geometry (or test of the
GR Kerr solution) would 
require a high accuracy, challenging the abilities of present and future planned detectors. For example, LISA 
observations are expected to be able to produce accurate measurements of the first 3-5 multipole moments of the MBH 
(see~\cite{Poisson:1996tc,Ryan:1997hg,Barack:2006pq} for discussions regarding this problem), which is at
the boundary of DCSMG detectability.

%----------------------------------------------------------------------------------------------------------------
\subsection{Motion of Massive Particles}
\label{geodesics}

Let us now consider the equations of motion for massive test particles in
DCSMG.  The starting point is the action of such a particle moving along a worldline
$x^{\mu} = z^{\mu}(\lambda)$, where $\lambda$ parameterizes the trajectory. This 
action is given by (see e.g.~\cite{Poisson:2004lr})
\be
S^{}_{\MAT} = - m \int^{}_{\gamma} d\lambda\; \sqrt{-\met^{}_{\alpha \beta}(z) 
\dot{z}^{\alpha} \dot{z}^{\beta}} \,,
\ee
where $m$ is the particle mass and $\dot{z}^{\mu} = d z^{\mu}/d\lambda$ is the tangent
to the worldline $\gamma$.  The contribution to the matter stress-energy tensor can be obtained
by variation of $S^{}_{\MAT}$ with respect to the metric and yields
\be
T^{\alpha\beta}_{\MAT}(x^{\mu}) = m\int \frac{d\tau}{\sqrt{-g}}u^\alpha u^\beta
\updelta^{(4)}[\mb{x}-\mb{z}(\tau)] \,, \label{tmunu}
\ee
where $\tau$ denotes proper time (which is related to $\lambda$ via $d\tau = 
d\lambda\; \sqrt{-\met^{}_{\alpha \beta}(z) \dot{z}^{\alpha} \dot{z}^{\beta}}$), 
$u^{\mu} = dz^{\mu}/d\tau$ is the particle four-velocity ($\met^{}_{\mu\nu}u^{\mu}u^{\nu}=-1$),
$g$ denotes the metric determinant, and $\updelta^{(4)}$ is the four-dimensional
Dirac density ($\int d^4x\sqrt{-g}\,\updelta^{(4)}(\mb{x})=1$). 

The divergence of $T^{\alpha\beta}_{\MAT}$ is
\be
\nabla^{}_{\beta} T^{\alpha\beta}_{\MAT} = m\int \frac{d\tau}{\sqrt{-g}}
\frac{d^{2}z^{\alpha}}{d\tau^{2}}\updelta^{(4)}[\mb{x}-\mb{z}(\tau)] \,,
\ee
and, substituting this into the divergence of the field equations [Eq.~\eqref{EEs}], we obtain 
\be
- \left(\nabla^{\nu} \vartheta\right) \left(\beta \square \vartheta + \frac{\alpha}{4} \pont  \right) = 
m \int \frac{d\tau}{\sqrt{-g}} \frac{d^{2}z^{\alpha}}{d\tau^{2}}\updelta^{(4)}[\mb{x}-\mb{z}(\tau)]\,.
\label{div-EOM}
\ee
Since the left-hand side of this equation vanishes by virtue of the evolution equation [Eq.~\eqref{EOM}],
all point-like particles must follow geodesics:
\be
\frac{d^{2}z^{\alpha}}{d\tau^{2}} = 0\,.
\ee
Notice that this result is completely generic and does not rely on any approximation scheme.

Let us now consider the structure of the timelike geodesics of the spacetime described by the 
metric in Eq.~\eqref{sol:metric_elements}.  Since this metric is a small deformation of the Kerr solution, we can
follow the same steps that lead to the derivation of the Kerr geodesic equations in GR (see
e.g.~\cite{Chandrasekhar:1992bo}) in a form suitable for numerical implementation. The new metric is 
still stationary and axisymmetric, which means it possesses a timelike Killing vector with components 
$t^{\alpha} =  [1,0,0,0]$ and a spacelike Killing vector with components 
$\psi^{\alpha} = [0,0,0,1]$ respectively.   These two Killing vectors commute as
in the Kerr case, and we can use them to define two conserved quantities: the energy,  
$E = -t^{\alpha} u_{\alpha}$, and the angular momentum, $L = \phi^{\alpha} u_{\alpha}$.
In Boyer-Lindquist coordinates, the particle four-velocity has components 
$u^{\alpha} = [\dot{t},\dot{r},\dot{\theta},\dot{\phi}]$, where the overhead dots now stand for 
differentiation with respect to proper time $\tau$.   
We shall work here in reduced variables, where $E$ and $L$ are the energy and angular momentum per 
unit mass $m$.  

We can now use the energy and angular momentum definitions to solve for $\dot{t}$ and 
$\dot{\phi}$.  To second order in the slow-rotation approximation, we have 
\be
\dot{t} = \dot{t}^{}_{K}+ L \delta g_{\phi}^{\CS}\,,
\qquad
\dot{\phi} = \dot{\phi}^{}_{K} - E \delta g_{\phi}^{\CS}\,,
\label{geo1}
\ee
where $\dot{t}^{}_{K}$ and $\dot{\phi}^{}_{K}$ denote the corresponding expressions for Kerr:
\ba
\label{tdot-GR}
\rho^{2} \dot{t}_{K} &=& \left[-a \left(a E \sin^{2}{\theta} - L \right) + \left(r^{2} 
+ a^{2} \right) \frac{P}{\Delta} \right]\,,  \\
\label{phidot-GR}
\rho^{2} \dot{\phi}_{K} &=& \left[- \left(a E - \frac{L}{\sin^{2}{\theta}} \right) 
+ \frac{a P}{\Delta} \right]\,,
\ea
and where $P = E (r^2 + a^2) - a L$, while $\delta g_{\phi}^{\CS}$ denotes the CS correction:
\be
\label{deltag}
\delta g_{\phi}^{\CS} = \frac{\alpha^{2}}{\beta \kappa} \frac{a}{112 r^{8} f} 
\left(70 r^{2} + 120 r\,M + 189 M^{2} \right)\,. 
\ee
As before, we have here resumed the Kerr part of the equations, so that 
when we take $\alpha \to 0$ we recover the exact equations for the Kerr spacetime
for all $a$, although the expressions presented in this paper are only formally 
valid up to second order in $a/M$ and $\zeta$. 

Following closely the Kerr case, we look for a Killing tensor to construct an
additional constant of motion: the Carter constant.  We find that such a tensor
can be written in the same form as in the Kerr case, namely 
\be
\xi^{}_{\alpha \beta} = \Delta\, k^{}_{(\alpha} l^{}_{\beta)} + r^{2}\, \bar{\met}^{}_{\alpha \beta}\,,
\ee
where $k^{\alpha}$ and $l^{\alpha}$ are two null vectors given by
\ba
k^{\alpha} = \left[\frac{r^{2}+a^{2}}{\Delta},-1,0,\frac{a}{\Delta} - \delta g_{\phi}^{\CS}\right]\,,
\nonumber \\
l^{\alpha} =  \left[\frac{r^{2}+a^{2}}{\Delta},1,0,\frac{a}{\Delta} - \delta g_{\phi}^{\CS}\right]\,,
\ea
that is, they are modifications of the Kerr principal null directions . We have checked that they 
are also principal directions of the new metric up to the order considered here. One can show by direct 
evaluation that this tensor satisfies the tensor Killing equations, 
$\nabla^{}_{(\rho}\,\xi^{}_{\alpha \beta)} = 0$,  on the CS modified Kerr background if and 
only if $\delta g_{\phi}^{\CS}$ is given by  Eq.~\eqref{deltag}. In fact, the tensor Killing equations 
allow us to add a term of the form $C/r^{2}$ to Eq.~\eqref{deltag}, but the orthogonality relation 
$k^{\alpha} l_{\alpha} = 0$ forces $C = 0$. 

We can now define the Carter constant in the same way as in the Kerr case
\be
{\cal{Q}} = \xi^{}_{\alpha \beta} u^{\alpha} u^{\beta} -  \left(L - a E \right)^{2}\,.
\ee
Using this relation, together with Eqs.~\eqref{tdot-GR} and~\eqref{phidot-GR} and the normalization condition 
for timelike geodesics, $\met^{}_{\alpha \beta} u^{\alpha} u^{\beta} = -1$, we can solve for 
$\dot{r}$ and $\dot{\theta}$.  We find that the result can be written as
\be
\dot{r}^{2} = \dot{r}_{K}^{2} +  2 E L f \delta g_{\phi}^{\CS}\,, \qquad
\dot{\theta}^{2} = \dot{\theta}_{K}^{2}\,, \label{geo2}
\ee
where the part that corresponds to Kerr (without expanding in $a/M$) is
\ba
\label{rdot-GR}
\rho^{4}\, \dot{r}^{2}_{K} &=&  \left[ \left( r^{2} + a^{2} \right) E - a L \right]^{2} 
\nonumber \\
&-&  \Delta\; \left[Q + \left(a E - L \right)^{2} + r^{2} \right]\,, \\
\label{thetadot-GR}
\rho^{4}\, \dot{\theta}^{2}_{K} &=& Q - \cot^{2}{\theta} L^{2} 
- a^{2} \cos^{2}{\theta} \left(1 - E^{2} \right)\,.
\ea

To summarize, Eqs.~\eqref{geo1} and~\eqref{geo2} provide a set of four first-order (in derivatives
of $\tau$) decoupled geodesic equations for the CS modified Kerr background.  We note that these geodesic 
equations contain CS corrections except for the $\dot{\theta}$ equation, and that these corrections have 
essentially the same structure, in the sense that all of them are proportional to $\delta g_{\phi}^{\CS}$
[Eq.~\eqref{deltag}]. We remind the reader once more that the polar angle $\theta$ and its evolution equation
$\dot{\theta}$ is not to be confused with the CS scalar field $\vartheta$ and its time derivative $\dot{\vartheta}$.  

One can easily show that the location of the innermost stable circular orbit (ISCO) for equatorial orbits 
in the CS modified metric is given by~\cite{Yunes:2009hc}
\be
R^{}_{\mbox{\tiny ISCO}}=6M \mp \frac{4\sqrt{6}a}{3}-\frac{7a^2}{18 M}
         \pm \frac{77\sqrt{6}a}{5184} \frac{\alpha^{2}}{\beta \kappa M^{4}}\,, \label{RISCO}
\ee
where the upper (lower) signs correspond to co-rotating (counter-rotating) geodesics. Then, it appears 
that the CS correction works against the spin effects, which suggests that similar behavior will be present 
in the solution of the CS modified geodesic equations.

%----------------------------------------------------------------------------------------------------------------
\subsection{Frequencies of the Orbital Motion}

The Kerr metric allows bound and stable geodesic trajectories (orbits) to be associated or decomposed
in terms of three {\em fundamental} frequencies (see e.g.~\cite{Schmidt:2002qk,Drasco:2003ky}): 
$\Omega^{}_{r}$ characterizes the radial motion (from periapsis to apoapsis and back); 
$\Omega^{}_{\theta}$ characterizes the motion in the polar direction; and $\Omega^{}_{\phi}$ characterizes 
azimuthal motion.  These frequencies are important because precessional orbital effects are due to
mismatches between them and because they can be used to decompose, among other things, the gravitational
waveform in a Fourier expansion.  

Expressions for these frequencies in terms of quadratures have been obtained for Kerr in~\cite{Schmidt:2002qk},
and recently also in~\cite{Drasco:2003ky}, where a new time coordinate is employed (see~\cite{Mino:2003yg}): 
$\rho^{2} d/d\tau = d/d\lambda$. In terms of the time $\lambda$, the modified geodesic equations become
\ba
\frac{d t}{d \lambda} &=& T^{}_{K}(r,\theta) + T_{\CS}(r)\,,~~
\left(\frac{d r}{d \lambda} \right)^{2} = R^{}_{K}(r) + R_{\CS}(r)\,,
\nonumber \\
\frac{d \phi}{d \lambda} &=& \Phi^{}_{K}(r,\theta) + \Phi_{\CS}(r)\,,~
\left(\frac{d \theta}{d \lambda} \right)^{2} = \Theta^{}_{K}(\theta)\,,
\ea
where $(T^{}_{K},R^{}_{K},\Theta^{}_{K},\Phi^{}_{K})$ are given by the right-hand sides of 
Eqs.~\eqref{tdot-GR},~\eqref{rdot-GR},~\eqref{thetadot-GR}, and~\eqref{phidot-GR} respectively, 
while $(T^{}_{\CS},R^{}_{\CS},\Phi^{}_{\CS})$ are quantities proportional to $\delta g_{\phi}^{\CS}$. 
Note that the CS corrections only depend on $r$, which is a consequence of the linearization in $a/M$. 

The fundamental frequencies of the radial and polar motions associated with the $\lambda$ time are
\be
{\cal{Y}}^{}_{r} = \frac{2\pi}{\Lambda^{}_{r}} \sim  
\frac{2 \pi}{\Lambda_{r}^{K}}\left(1 - \frac{\Lambda_{r}^{\CS}}{\Lambda_{r}^{K}}\right)\,,~~ 
{\cal{Y}}^{}_{\theta} =  \frac{2\pi}{\Lambda^{}_{\theta}} = \frac{2\pi}{\Lambda_{\theta}^{K}}\,,
\ee
where the periods, denoted by $\Lambda$, are given by 
\ba
\Lambda_{r}^{K} & = & 2 \int^{r^{}_{\APO}}_{r^{}_{\PERI}} \frac{dr}{\sqrt{R^{}_{K}}}\,, ~
\Lambda_{r}^{\CS} = - \int^{r^{}_{\APO}}_{r^{}_{\PERI}} dr \frac{R_{\CS}}{R_{K}^{3/2}}\,, \\
\Lambda_{\theta}^{K} & = & 2 \int^{\pi-\theta^{}_{\MIN}}_{\theta^{}_{\MIN}} 
\frac{d\theta}{\sqrt{\Theta^{}_{K}}} = 4\int^{\pi/2}_{\theta^{}_{\MIN}} 
\frac{d\theta}{\sqrt{\Theta^{}_{K}}}  \,,
\ea 
where $r^{}_{\APO}$ and $r^{}_{\PERI}$ are the apocenter and pericenter values of $r$ respectively, 
and $\theta^{}_{\MIN}$ determines the interval in which $\theta$ oscillates, i.e. 
$(\theta^{}_{\MIN},\pi-\theta^{}_{\MIN})$.  The frequency associated with the 
azimuthal motion is given by~\cite{Drasco:2003ky}
\be
{\cal{Y}}^{}_{\phi} = \frac{1}{(2 \pi)^{2}} \int_{0}^{2 \pi} \!\!\!dw^{r} 
\int_{0}^{2 \pi} \!\!\!dw^{\theta}\; \Phi[r(w^{r}),\theta(w^{\theta})]\,,
\ee
and where $w^{r,\theta} = {\cal{Y}}^{}_{r,\theta} \lambda$ are the associated angle variables.
Therefore, only ${\cal{Y}}^{}_{r}$ and ${\cal{Y}}^{}_{\phi}$ change with respect to GR.

The expressions for the fundamental frequencies with respect to the coordinate time $t$,
$\Omega^{}_{r,\theta,\phi}$ are more involved since they are given by $\Omega^{}_{r,\theta,\phi} = 
{\cal{Y}}^{}_{r,\theta,\phi}/\Gamma$, and $\Gamma$ is~\cite{Drasco:2003ky}
\be
\Gamma = \frac{1}{(2 \pi)^{2}} \int_{0}^{2 \pi} \!\!\!dw^{r} \int_{0}^{2 \pi}\!\!\! dw^{\theta} \; 
T[r(w^{r}),\theta(w^{\theta})]\,.
\ee
Inverting this relation cannot be easily done in close form, without previously solving the integrals 
numerically (see eg.~\cite{Schmidt:2002qk} for an analytical inversion in terms of incomplete elliptical integrals). 
We immediately see, however, that since $T$ is CS corrected, so will be $\Gamma$, 
leading to CS corrections in all the frequencies associated with the coordinate time $t$.

Since the three physical fundamental frequencies of the Kerr spacetime are modified in CS modified gravity, given a 
detection of quasinormal modes from a Kerr-like object, the inference of system parameter could lead to 
a systematic error due to the bias of assuming GR is the correct underlying theory~\cite{bias}. 
Conversely, the detection of such modes with LISA may yield an interesting constraint on dynamical 
CS modified gravity.

%%%%%%%%%%%%%%%%%%%%%%%%%%%%%%%%%%%%%%
\section{Gravitational Wave Generation and the Semi-Relativistic Approximation}
\label{h-quad}

In this section we look at the structure of the modified field equations of DCSMG to study 
how gravitational-wave generation formulae change with respect to the GR ones.  
We also describe the semi-relativistic approximation for the description
of IMRI/EMRI systems and the details of the implementation that we use in this paper.

%----------------------------------------------------------------------------------------------------
\subsection{GW generation in DCSMG}

Taking into account the extreme mass ratios involved, the systems under considerations can be
well described using perturbation theory around the background quantities given in 
Sec.~\ref{backgroundquantities}. In that section, we showed that these background quantities contain
a GR contribution plus perturbations in the CS coupling parameter $\zeta$ and in the
black hole rotation parameter $a/M$. Similarly, here we shall expand 
the perturbations generated by the SCO (treated here as a particle) orbiting the MBH also in powers 
of a third parameter: the mass ratio of the system $\mu = m/M$.  

Neglecting the slow-rotation approximation for the time being, the metric and the CS scalar field
can be expanded as follows:
\ba
\met^{}_{\mu\nu} & = &\bar\met^{}_{\mu\nu} + \sum^{}_{n=1,m=0} \mu^{n}\zeta^{m} h^{(n,m)}_{\mu\nu} \nn \\
& = & \bar\met^{}_{\mu\nu} + \mu h^{(1,0)}_{\mu\nu} + \mu\zeta h^{(1,1)}_{\mu\nu} + \ldots\,,
\ea
\ba
\vartheta & = & \bar\vartheta + \sum^{}_{n=1,m=0} \mu^{n}\zeta^{m} \vartheta^{(n,m)} \nn \\ 
& = &  \bar\vartheta + \mu\vartheta^{(1,0)} + \mu\zeta\vartheta^{(1,1)} + \ldots \,,
\ea
where $(\bar\met^{}_{\mu\nu},\bar\vartheta)$ are given in Eqs.~\eqref{sol:metric_elements} and~\eqref{back-theta}
and $h^{(n,m)}_{\mu \nu}$ and $\vartheta^{(n,m)}$ now stand for the perturbations of ${\cal{O}}(\mu^{n},\zeta^{m})$ 
(alternatively, one can think of $\mu$ as playing the role of $\epsilon$ in Sec.~\ref{smallcouplingapprox}).

The equation for the metric perturbation $h^{(1,0)}_{\mu\nu}$, in the Lorenz gauge [Eq.~\eqref{lorenzgauge}], is:
\be
\mu\, \{ \bar\square h^{(1,0)}_{\mu\nu} + 2\,\bar{R}^{\rho}{}^{}_{\mu}{}^{\sigma}{}^{}_{\nu} 
h^{(1,0)}_{\rho\sigma} \} = - \frac{1}{\kappa} T^{\MAT}_{\mu\nu} \,, \label{1stmetpert}
\ee
where there are no contributions from the CS scalar field at this order of approximation, and thus, 
this is a purely GR GW generation equation.
Similarly, the equation for $\vartheta^{(1,0)}$ is:
\be
\bar\square \vartheta^{(1,0)} =  h^{(1,0)\mu\nu}\,\bar\nabla^{}_{\mu}\bar\nabla^{}_{\nu}\bar\vartheta
+\bar\met^{\mu\nu}\, \oone \Gamma^{\rho}_{\mu\nu}\bar\nabla^{}_{\rho}\bar\vartheta 
-  \frac{\alpha}{4\beta}\,\oone\!\left(\pont\right)\,, \label{1stfieldpert}
\ee
where $\oone\Gamma^{\rho}_{\mu\nu}$ and $\oone\!\left(\pont\right)$ denote the first-order
perturbations of the Christoffel symbols and the Pontryagin density respectively, 
about $\bar\met^{}_{\mu\nu}$ and evaluated at $h^{(1,0)}_{\mu\nu}$. 
That is, $\vartheta^{(1,0)}$ is generated by the metric perturbations
$h^{(1,0)}_{\mu\nu}$ which, in turn, are generated by the motion of the SCO around the MBH.  
The first two terms on the right-hand side of Eq.~\eqref{1stfieldpert} are proportional to
the spin parameter $a/M$. This is not the case for the third term, since
this is non-zero for a Schwarzschild BH~\cite{Yunes:2007ss}, which implies that $\vartheta^{(1,0)}$
will contain terms proportional to $a$ but also terms that are $a$-independent.

The next correction to the metric perturbations in the small coupling approximation is 
$h^{(1,1)}_{\mu\nu}$, which satisfies an equation with the following structure
\ba
& & \zeta\left\{\bar\square h^{(1,1)}_{\mu\nu} + 2\,\bar{R}^{\rho}{}^{}_{\mu}{}^{\sigma}{}^{}_{\nu} 
h^{(1,1)}_{\rho\sigma}\right\} =  \nn  \\
& &\qquad \frac{2\alpha}{\kappa}\left\{ C^{}_{\mu\nu}[\vartheta^{(1,0)},\bar\met^{}_{\rho\sigma}] 
+ \oone C^{}_{\mu\nu}[\bar\vartheta,h^{(1,0)}_{\mu\nu}]\right\} \nn \\
& & \qquad - 2 \frac{\beta}{\kappa} \left\{ \bar\nabla^{}_{(\mu}\bar\vartheta
\bar\nabla^{}_{\nu)}\vartheta^{(1,0)}-\frac{1}{2}\bar\met^{}_{\mu\nu}\bar\nabla^{\sigma}
\bar\vartheta\bar\nabla^{}_{\sigma}\vartheta^{(1,0)} \right. \nn \\
& & \qquad \left. -\frac{1}{4} h^{(1,0)}_{\mu\nu}\,
\bar\nabla^{\sigma}\bar\vartheta\bar\nabla^{}_{\sigma}\bar\vartheta\right\}\,, \label{eqgwh11}
\ea
where $\oone C^{}_{\mu\nu}$ is the first-order perturbation of the C-tensor when only 
the metric is perturbed [see Appendix~\ref{2ndswa} for explicit formulae of these quantities].
Note that the terms on the right-hand side of Eq.~\eqref{eqgwh11} depend only on the background (over-head barred) quantities and
the $(1,0)$ perturbations, which are determined by Eqs.~\eqref{1stmetpert} and~\eqref{1stfieldpert}.

The leading-order CS correction to the GW generation formalism is then determined by Eq.~\eqref{eqgwh11},
so let us now analyze its structure. The first term in this equation is 
$(2\alpha/\kappa)C^{}_{\mu\nu}[\vartheta^{(1,0)},\bar\met^{}_{\rho\sigma}]$, 
and it contains two pieces:
\ba
\label{first-first}
&& (2\alpha/\kappa)(\bar\nabla^{}_{\sigma} \vartheta^{(1,0)})
\bar\epsilon^{\,\sigma \delta \alpha(\mu}\bar\nabla^{}_{\alpha}\bar{R}^{\nu)}{}^{}_{\delta}\,,
 \\
&& (2\alpha/\kappa)(\bar\nabla^{}_{\sigma}\bar\nabla^{}_{\delta}\vartheta^{(1,0)}) 
{\,^\ast\!}\bar{R}^{\delta (\mu \nu)\sigma}\,. \label{secondpiece}
\ea
Since the background is Kerr plus terms of order $\zeta$, Eq.~\eqref{first-first} is actually at least of
${\cal{O}}(\zeta^{2} \, \mu \, a/M)$, and we can then ignore it to compute $h^{(1,1)}_{\mu\nu}$. 
The second piece [Eq.~\eqref{secondpiece}] is actually the only one in Eq.~\eqref{eqgwh11} 
that is not proportional to the spin parameter, making it of ${\cal{O}}(\zeta \; \mu)$, 
since the dual to the Riemann of the Schwarzschild metric is non-vanishing. 
The last term in Eq.~\eqref{eqgwh11} is at least quadratic in $a/M$, making it of ${\cal{O}}(\zeta \; \mu \; a^2/M^{2})$, 
since $\bar{\vartheta}$ is linear in the spin parameter. 
All remaining terms are at least linear in the spin parameter and of ${\cal{O}}(\zeta \; \mu \; a/M)$, 
since they are all proportional to derivatives of $\bar\vartheta$, 
which in turn is proportional to $a/M$ [see Eq.~\eqref{back-theta}].

The structure of Eq.~\eqref{eqgwh11} thus reveals that if one were to employ the slow-rotation approximation, 
as in the computation of the MBH metric~\cite{Yunes:2009hc} (see Sec.~\ref{backgroundquantities}), 
the leading-order evolution equation for the metric perturbation becomes
\be
\zeta\left\{\bar\square h^{(1,1)}_{\mu\nu} + 2\,\bar{R}^{\rho}{}^{}_{\mu}{}^{\sigma}{}^{}_{\nu} 
h^{(1,1)}_{\rho\sigma}\right\} = 
 \frac{2\alpha}{\kappa} (\bar\nabla^{}_{\sigma}\bar\nabla^{}_{\delta}\vartheta^{(1,0)}) 
{\,^\ast\!}\bar{R}^{\delta (\mu \nu)\sigma}\,. \label{eqgwh11-leading}
\ee
Once this equation is solved for $h^{(1,1)}_{\mu\nu}$, one could use it to solve the evolution equation for $\vartheta^{(1,1)}$, 
which we have not presented here. In this way, one could proceed with the boot-strapping method 
described in~\cite{Yunes:2009hc} and in Sec.~\ref{lin-th} to build higher-order perturbations in the metric 
and the CS scalar field.  

In summary, GW emission in DCSMG is determined by the following equations. To first order, the metric perturbation
is generated by the SCO through the GR relation in Eq.~\eqref{1stmetpert}, leading to $h^{(1,0)}_{\mu\nu} = {\cal O}({\mu})$.
To second order, the metric perturbation contains two contributions: a term $h_{\mu \nu}^{(2,0)} = {\cal O}({\mu}^{2})$ and
a CS modification to the first-order term $h_{\mu \nu}^{(1,1)} = {\cal O}({\mu}\zeta)$. The latter can be obtained by solving 
Eq.~\eqref{eqgwh11-leading}, which requires knowledge of the leading-order perturbation to the CS scalar due to the SCO, 
$\vartheta^{(1,0)}$, which in turn is determined by Eq.~\eqref{1stfieldpert} once one has solved for $h_{\mu \nu}^{(1,0)}$ through 
Eq.~\eqref{1stmetpert}.  In this paper we shall only employ the first-order terms, $h^{(1,0)}_{\mu\nu}$, to model GWs, and thus
modifications to the waveforms arise exclusively due to corrections to the trajectories.

A proper treatment of the EMRI problem would also require the use of the first-order perturbations to 
estimate backreaction, self-force effects on the SCO trajectory (see e.g.~\cite{Poisson:2004lr}),
which involves calculations that have {\emph{not}} been carried out in full generality even in pure GR.  
Only through the adiabatic approximation has there been studies of such back-reaction for spinning 
MBHs~\cite{Apostolatos:1993nu,Cutler:1994pb,Kennefick:1995za,Kennefick:1998ab,Mino:2003yg,Hughes:2005qb,Drasco:2005kz,Sago:2005gd,Ganz:2007rf,Fujita:2009rr}, 
where one assumes that only GW dissipation is important.
The purposes of this paper, however, is to elucidate the main differences between
GR and DCSMG dynamics for IMRI/EMRI systems, and thus, calculations with back-reaction are far too complex. 
Instead, we will resort to the so-called semi-relativistic approximation, which we describe in the next
subsection.

%-----------------------------------------------------------
\subsection{The Semi-Relativistic Approximation}\label{semirelativistic}

First introduced by Ruffini and Sasaki~\cite{Ruffini:1981rs}, this approximation assumes that the SCO
moves along geodesics of the MBH background geometry and, more importantly, that the SCO emits GWs
as if it were moving in flat space.  These are two strong assumptions but they give a good qualitative
picture of the dynamics.  In fact, it has been recently shown~\cite{Babak:2006uv} that in combination with some
post-Newtonian prescriptions for radiation reaction, semi-relativistic like calculations lead to waveforms 
with high overlaps with Teukolsky-based waveforms (waveforms were the GW fluxes for
the adiabatic approximation are calculated from solutions to the Teukolsky equation for perturbations of
a Kerr background).  The geodesic motion assumption is justified for EMRIs because the more extreme the mass ratio, the 
closer one is to the point-particle limit, and hence the closer the motion is to geodesic. The flat-space assumption 
allows us to neglect the curvature of the background in the generation and propagation of GWs from the
SCO to the observer in the radiation zone.  

Following the work of Ruffini and Sasaki~\cite{Ruffini:1981rs}, the implementation of the semi-relativistic 
approximation is as follows. Once the geodesic equations have been solved for the orbital trajectories, 
the metric perturbation is obtained from Eq.~\eqref{1stmetpert},
but assuming that the background metric $\bar\met^{}_{\mu\nu}$ is the flat spacetime metric. 
In this way, the solution to the perturbative equation is given by the gravitational Lienard-Wiechert
potentials (see e.g.~\cite{Barut:1980ao,Jackson:99jd} for the derivation of the electromagnetic version 
of these potentials), which contain all the information necessary to reconstruct the gravitational field generated
by the SCO.

In this paper, we shall employ a formulation of the semi-relativistic approximation in which the first assumption
(geodesic motion with respect to the MBH background geometry) is still employed, but the
second one is slightly modified.  More precisely, we shall still assume that GWs are emitted as if in flat space
but, following~\cite{Babak:2006uv,Yunes:2008gb}, we solve the perturbative equations using a multipolar expansion, 
which is determined by the source multipole moments and which is valid for objects with arbitrarily 
strong internal gravity (see~\cite{Thorne:1980rm} for a detailed discussion of these expansions). 

The multipolar treatment of gravitational radiation is truly a slow-motion approximation as 
the series is truncated at a finite multipole order.  For a compact-body binary system, the error  
scales with a high-power of the orbital velocity, which for EMRIs can typically be larger than half the
speed of light, and may lead to the loss of certain relativistic features in the waveforms (see eg.~\cite{Yunes:2008tw}
in the context of post-Newtonian theory and EMRIs). 
In spite of these drawbacks, the semi-relativistic approximation has been shown to capture the correct 
qualitative behavior of EMRI orbits and waveforms, and thus, it will suffice to present the main 
features of the CS modifications to GW generation. 

Another possibility to estimate the GW emission is
to use the weak-gravity/fast-motion formula derived by Press~\cite{Press:1977ps}, whose implementation
requires the same information as the multipolar formula truncated after the mass octopole and current
quadrupole.  This formula has been used in~\cite{Babak:2006uv}, where it has been shown to provide
accurate results as compared to Teukolsky-based waveforms. We shall not employ this formalism here, however, 
choosing to work instead with the standard multipolar decomposition.

We have shown that up to order ${\cal O}(\mu\zeta)$ the perturbative GW generation equations for IMRI/EMRI systems 
have the same form as in GR. Any difference in the waveforms will then arise due to differences in the geodesic equations due to
the modified Kerr background. In the transverse-traceless (TT) gauge, 
defined by Eqs.~\eqref{lorenzgauge} and~\eqref{tracelesscondition} and in which the dynamical degrees of freedom have been 
isolated, the metric perturbations describing the GWs emitted by an isolated system are given by~\cite{Thorne:1980rm}
\ba
h_{ij}^{\TT} &=& \left[\;\sum_{\ell=2}^{\infty} \frac{4}{\ell!\, r}\, I^{(\ell)}_{ijA^{}_{\ell-2}}\!(t^{}_{ret})\, 
N^{}_{A^{}_{\ell-2}} \right. \nonumber \\
&+& \left. \sum_{\ell=2}^{\infty} \frac{8\,\ell}{(\ell+1)!\,r}\, \varepsilon^{}_{kl(i}\, J^{(\ell)}_{j)k 
A^{}_{\ell-2}}\!(t^{}_{ret})\,n^{}_{l} N^{}_{pA^{}_{\ell-2}}\right]^{\TT}\!\!\!\!\!\!\!\!\!\,, \label{GR-sol}
\ea
where we have adopted the multi-index notation of~\cite{Thorne:1980rm}: $A^{}_{\ell}$ is a multi-index (sequence
of $\ell$ indices: $a^{}_{1}\cdots a^{}_{\ell}$) so that $N^{}_{A^{}_{\ell}} = n_{a^{}_{1}} \cdots n^{}_{a^{}_{\ell}}$ is 
the product of $\ell$ unit vectors $n^{i} = x^{i}/r$ that point in the direction from the source toward the observer, 
with $r = (\delta_{ij} x^{i} x^{j})^{1/2}$ is the flat-space distance from the source to the observer; 
there is summation over repeated indices independently of their location;
$\varepsilon^{}_{ijk}$ is the $3$-dimensional flat-space Levi-Civita tensor. 

The mass and current multipole moments of the source, $I^{}_{A^{}_{\ell}}$ and $J^{}_{A^{}_{\ell}}$,  are given 
by~\cite{Thorne:1980rm}
\ba
I^{}_{A^{}_{\ell}} &=& \left[ \int d^{3}x \; \rho \; x^{}_{A^{}_{\ell}} \right]^{\STF}\,, 
\nonumber \\
J^{}_{A^{}_{\ell}} &=& \left[ \int d^{3}x \; \rho \; (\varepsilon^{}_{a^{}_{\ell}jk} x^{}_{i} v^{}_{k})\,
x^{}_{A^{}_{\ell-1}} \right]^{\STF}\,, \label{masscurrentmultipoles}
\ea
where $x^{}_{A^{}_{\ell}} = x^{}_{a^{}_{1}}\cdots x^{}_{a^{}_{\ell}}$ and $\rho$ is the mass density of the 
source, which in our case corresponds to the mass density of the SCO, given by
\be
\rho(t,x^{i}) =  m\, \updelta^{(3)}[x^{i}-z^{i}(t)]\,,
\ee
where $\updelta^{(3)}$ denotes the Dirac-delta distribution, $z^{i}(t)$ is the spatial trajectory of the SCO, 
and $v^{i} = dz^{i}/dt$ its spatial velocity with respect to the coordinate time $t$.  
In Eq.~\eqref{GR-sol} the mass and current multipole moments
have to be evaluated at the retarded time $t^{}_{ret} = t - r$ and the superscript $(\ell)$ denotes
their $\ell$-th time derivative

The projectors $\STF$ in Eqs.~\eqref{masscurrentmultipoles} and $\TT$ in Eq.~\eqref{GR-sol} 
are symmetric/trace-free and transverse/traceless operators.  The latter is obtained by means of the
projector orthogonal to $n^{i}$, namely $P^{i}_{j} = \delta^{i}_{j}-n^{i}n^{}_{j}$, in the following way
\be
\left[A^{}_{ij}\right]^{\TT} = P_{i}^{k}P_{j}^{l}A^{}_{kl} -\frac{1}{2}P^{}_{ij}P^{kl}A^{}_{kl}\,.
\ee
An orthonormal triad, $\{n^{i},p^{i},q^{i}\}$, with respect to the spatial flat-space metric and associated
with $n^{i}$, can always be constructed to build the polarization tensors via
$\varepsilon^{+}_{ij} = p^{}_ip^{}_j - q^{}_iq^{}_j$ and $\varepsilon^{\times}_{ij} = 2p^{}_{(i}q^{}_{j)}$.

With these tensor projectors and Eq.~\eqref{GR-sol} we can obtain the two independent GW polarizations as follows:
\be
h^{}_+ = \frac{1}{2}\varepsilon^{+ij} h^{\TT}_{ij} \,, \qquad
h^{}_{\times} = \frac{1}{2}\varepsilon^{\times ij} h^{\TT}_{ij}\,,
\ee
or equivalently
\be
h^{\TT}_{ij} = h^{}_+\, \varepsilon^{+}_{ij} + h^{}_{\times}\,\varepsilon^{\times}_{ij}\,.
\ee
In this paper we use this multipolar decomposition to compute the GW emission including terms up to 
the mass octopole and current quadrupole moment. 

Another important question for the implementation of this version of the semi-relativistic
approximation is the choice of coordinates.  More specifically, the equations
of motion (geodesics) are written in Boyer-Lindquist coordinates (see Sec.~\ref{geodesics}), whereas
the formulae given in this section require Cartesian coordinates.  Thus, one must decide how to
construct Cartesian coordinates that cover the motion of the SCO and at the same time the observer
in the radiation zone.  
One possibility would be to rewrite the background MBH metric and the equations
of motion in Cartesian-like coordinates such that at infinity the metric becomes the flat spacetime 
metric in Cartesian coordinates.  This choice was made in~\cite{Yunes:2008gb} for the study
of extreme-mass ratio gravitational-wave bursts. A different choice is to identify 
Boyer-Lindquist coordinates $(r,\theta,\phi)$ with flat-space spherical coordinates and then
introduce Cartesian coordinates through the familiar relations
\be 
x=r\sin\theta\cos\phi\,,~~
y=r\sin\theta\sin\phi\,,~~ 
z=r\cos\theta\,. \label{cartesianlike}
\ee
This was the choice made in~\cite{Babak:2006uv} for the construction
of EMRI {\em kludge} waveforms.  Numerical experiments~\cite{carlosunpub} show that indeed different
choices of Cartesian coordinates produce different waveforms.  However, differences are significant
only for relativistic orbits, and for those, these differences appear dominantly in the amplitudes but not in the phase. 
In this paper we shall employ the second choice of Cartesian coordinates for the computation of the gravitational
waveforms.

%%%%%%%%%%%%%%%%%%%%%%%%%%%%%%%%%%%%%%
\section{Gravitational Waves from IMRI/EMRI systems in DCSMG}
\label{semi-class}

In this section we study numerically the evolution of IMRI/EMRIs and their GW emission
in the semi-relativistic approximation and in the context of the DCSMG theory.
We describe the numerical implementation and apply it to a number of test systems,
presenting the trajectories and the waveforms associated with them.

%-------------------------------------------
\subsection{Numerical Evolution and Test Systems}
\label{test-sys}

In the semi-relativistic approximation the numerical calculations can be divided into two
stages.  First, the computation of the SCO trajectory around the MBH and, second, the
computation of the gravitational waveforms given the trajectory.  
The starting point for the computation of the trajectory, namely $[t(\tau),r(\tau),\theta(\tau),\phi(\tau)]$ 
are the geodesic equations [Eqs.~\eqref{geo1} and~\eqref{geo2}].  Clearly, when the CS parameter
$\xi$ is small the dynamics will be close to GR.  Since there are
already bounds on this parameter from binary pulsar tests~\cite{Yunes:2009hc}, albeit weak, 
we shall be here interested only in small CS deformation of GR EMRI/IMRI waveforms.

As in GR, for bound orbits the equations for $r(\tau)$ and $\theta(\tau)$ present turning points which 
are difficult to treat numerically.  This can be avoided by introducing the following alternative
variables (see e.g.~\cite{Drasco:2003ky} for details):
\be
r = \frac{p\,M}{1+e\cos\psi}\,,~~
\cos^{2}{\theta} = \cos^{2}({\theta^{}_{\MIN}}) \cos^{2}({\chi})\,, \label{rzangles}
\ee
where $p$ is the {\em semi-latus rectum}, $e$ is the eccentricity, and as we mention before
$\theta \in (\theta^{}_{\MIN},\pi-\theta^{}_{\MIN})$. The geodesic equations then become evolution
equations for the variables $(\psi,\chi,\phi)$ with respect to Boyer-Lindquist time $t$, where one uses
the $dt/d\tau$ equation [Eq.~\eqref{geo1}] to transform from proper to coordinate time.  
The form of the ordinary differential equations (ODEs) for 
$[\psi(t),\chi(t),\phi(t)]$ in GR can be found from the developments presented, for instance in~\cite{Drasco:2003ky}.  

From a numerical standpoint,  ODE solvers for $[\psi(t),\chi(t),\phi(t)]$ require knowledge of the turning points, 
which will be here CS modified. In particular, the radial turning points are modified, since the ODE for 
$\psi(t)$ is corrected by new CS terms that arise in the $dr/d\tau$ equation.  In contrast, 
the polar turning points are not modified, since these arise from the $d\theta/d\tau$ equation, which is 
not CS corrected. All of this suggests that prescribing a set of initial constants of motion $(E,L,Q)$ leads
to three orbital parameters $(p,e,\iota)$, which are not identical to the ones that one would obtain in GR. 

Let us define these orbital parameters more carefully, since they play an important role in our calculations.
The semi-latus rectum and the eccentricity parameter, $p$ and $e$, are defined such that 
the apocenter $r^{}_{\APO}$ and pericenter $r^{}_{\PERI}$ (two of the turning points of the equation for $dr/d\tau$)
have the familiar Newtonian expression
\be
r^{}_{\APO} = \frac{p\,M}{1-e}\,,~~ r^{}_{\PERI} = \frac{p\,M}{1+e}\,.
\ee
The third orbital parameter, the orbital inclination with respect to the equatorial plane $\iota$, is defined as
$\tan\iota = Q/L$, but it is sometimes more convenient to describe it in terms
of the angle $\theta^{}_{\INC} = \pi/2 - \mbox{\rm sign}(L)\theta^{}_{\MIN}$, which is directly related to the
turning points of $\theta(\tau)$.  Clearly, the functional form of $(p,e,\theta^{}_{\INC})$
in terms of $(E,L,Q)$ differs in GR and in DCSMG, and thus, if one prescribes a set of constants of motion
$(E_{0},L_{0},Q_{0})$ in GR the resultant orbit will not possess the same orbital parameter as its DCSMG counterpart.  
In the results presented below, we compare trajectories that have the {\emph{same}} orbital parameters, 
$(p,e,\theta^{}_{\INC})$, instead of comparing trajectories with the same constants of motion $(E,L,Q)$.  

Regarding the numerical integration of the ODEs for $[\psi(t),\chi(t),\phi(t)]$, we use  
the Bulirsh-Stoer extrapolation method~\cite{Bulirsch:1966bs} as the evolution
algorithm (details on this method and its implementation can be found in~\cite{Press:1992nr,Stoer:1993sb}).
Our numerical code is complemented with a number of routines to construct Boyer-Lindquist and Cartesian-like
coordinates and their associated time derivatives.

Waveforms are computed with the multipolar expansion of Sec.~\ref{semirelativistic}, up to the mass octupole 
and current quadrupole, which requires as input the trajectory $x^{i}(t)$,
the spatial velocity $v^{i}(t)$, the spatial acceleration $a^{i}(t)$, and its time derivative, the {\em jerk}, 
$j^{i}(t) = da^{i}(t)/dt$.  These time derivatives are computed using a finite differences differentiation rule 
with nine points.  The error in the computation of the jerk scales with the time step, $\Delta t$, as $(\Delta t)^{8}$, 
well within the accuracy range that we need for our calculations.

We have here evolved the geodesic equations for a total time of $T = 5 \times 10^{5} M$, obtaining on
the order of $10^{3}$-$10^{4}$ cycles.  The type of geodesic is prescribed in terms of the following 
orbital parameters: the pericenter {\em distance} $r^{}_{\PERI}$, the eccentricity $e$, and the inclination 
angle $\theta^{}_{\INC}$.  
We implemented a double-bisection algorithm to guarantee that the turning points in both the GR evolutions 
and the DCSMG ones correspond to orbits with the same orbital parameters $(p,e,\theta^{}_{\INC})$, up to an 
accuracy of one part in $10^{14}$.  

The MBH background metric of Eq.~\eqref{sol:metric_elements} is fully determined by the choice of the mass $M$ 
and spin parameter $a = J/M$, where $J$ is the spin angular momentum, plus the ratio $\xi$.
The SCO is characterized by its mass $m$, which then also defines the mass ratio 
$\mu = m/M$.  As we have mentioned, the type of orbit is fully determined by the choice of orbital parameters
$(r^{}_{\PERI},e,\theta^{}_{\INC})$, while we still need to fully determine the initial conditions of the motion
by prescribing $(\psi^{}_{o},\chi^{}_{o},\phi^{}_{o}) = (\psi(t^{}_{o}),\chi(t^{}_{o}),\phi(t^{}_{o}))$, where
$t^{}_{o}$ is the initial time for the numerical evolution of our system of ODEs.  We shall here always set 
$\beta = \alpha$, $M=M^{}_{\bullet}$ where $M^{}_{\bullet} = 4.5 \times 10^{6} M_{\odot}$ is approximately the mass 
of the presumable MBH at the center of the Milky Way~\cite{Gillessen:2009mw}.  For the mass of the SCO we
choose $m=35 M_{\odot}$, which leads to the following mass ratio: $\mu \sim 7.8 \times 10^{-6}$.

The particular {\em test} systems that we consider in this paper are defined as follows:
\begin{itemize}
\item[(A)\,:] $a=0.1\,M$ and $\xi = 0.1 M^{4}$, with orbital parameters 
              $(r^{}_{\PERI},e,\theta^{}_{\INC}) = (12\,M, 0.2, 0.1)\,$, 
\item[(B)\,:] $a=0.2\,M$ and $\xi = 0.2 M^{4}$, with orbital parameters 
              $(r^{}_{\PERI},e,\theta^{}_{\INC}) = (8\,M, 0.4, 0.2)\,$, 
\item[(C)\,:] $a=0.4\,M$ and $\xi = 0.4 M^{4}$, with orbital parameters 
              $(r^{}_{\PERI},e,\theta^{}_{\INC}) = (6\,M, 0.6, 0.3)\,$,  
\end{itemize}
while the waveforms are measured by observers located on the $z$-axis at a 
distance of approximately $8\,$kpc (roughly the distance from the Solar System to the center of our galaxy).  
Clearly, the test systems have been ordered from the least relativistic to the most relativistic.  
We are forced to choose $a$ and $\zeta$ such that the slow-rotation approximation $a/M \ll 1$ and the 
small-coupling approximation $\zeta \ll 1$ used for constructing the MBH background~\cite{Yunes:2009hc} 
are satisfied.  In this sense, the error in the MBH background metric scales as 
$\zeta (a^{3}/M^{3})$, which for systems $(A)$, $(B)$, and $(C)$ implies errors in the MBH
background of ${\cal{O}}(10^{-4})$, ${\cal{O}}(10^{-3})$ and ${\cal{O}}(10^{-2})$ respectively, 
relative to the true DCSMG waveforms for an exact, spinning MBH background.

%------------------------------------------------------------------------------------------------
\subsection{Orbital Trajectories}

\begin{figure}[htp]
\begin{center}
\includegraphics[scale=0.8,clip=true]{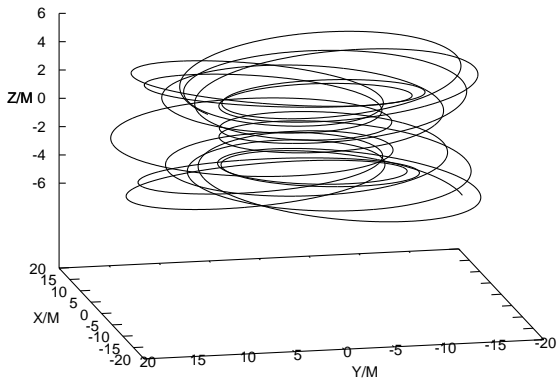}
\caption{\label{3dorb} Three-dimensional depiction of the last cycles of system B before the evolution was stopped.}
\end{center}
\end{figure}

The test systems described above present rather similar orbital behavior. Generically, there is a stage of 
zoom-whirl, where the particle spends several cycles close to the pericenter radius, $r^{}_{\PERI}$, followed by 
a stage where it orbits at a larger radius $r$, close to $r^{}_{\APO}$. Moreover, there is generically both 
in-plane and out-of-the-plane precession.  The orbit produced by test system $(B)$ in GR is shown in 
Fig.~\ref{3dorb}, where we have plotted only the last time interval of duration $17500 M$ of the geodesic evolution.
From this figure we can observe the generic Lense-Thirring precession as well as the two stages described above. 

%%%%%%%%%%
%%%%%%%%%%

\begin{figure}[htp]
\begin{center}
\includegraphics[scale=0.33,clip=true]{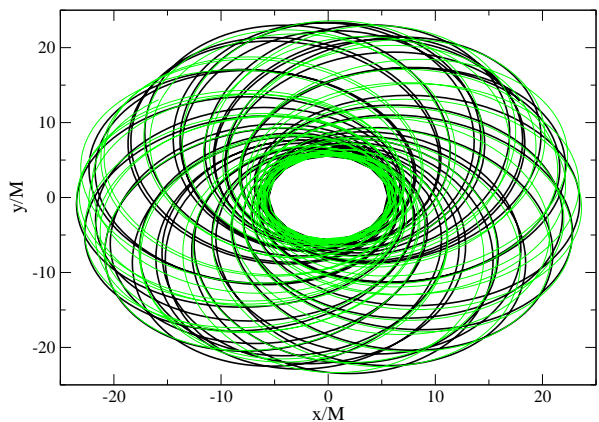}
\caption{\label{xy-plane} Two-dimensional projection onto the $x$-$y$ plane of system $(C)$'s orbital 
trajectories in a Kerr background (light gray) and in the CS corrected Kerr solution (black).}
\end{center}
\end{figure}

The CS correction to the background has a clear effect on the trajectories of the bodies.
Figure~\ref{xy-plane} plots the projection of the orbital trajectories for system C onto the $x$-$y$ plane for the CS 
background (black line) and the Kerr background (light gray line) and the last $17500 M$ of the geodesic evolution. 
Notice that the orbital trajectories have dephased significantly by the end of the evolution.   

%%%%%%%%%%
%%%%%%%%%%

\begin{figure}[htp]
\begin{center}
\includegraphics[scale=0.33,clip=true]{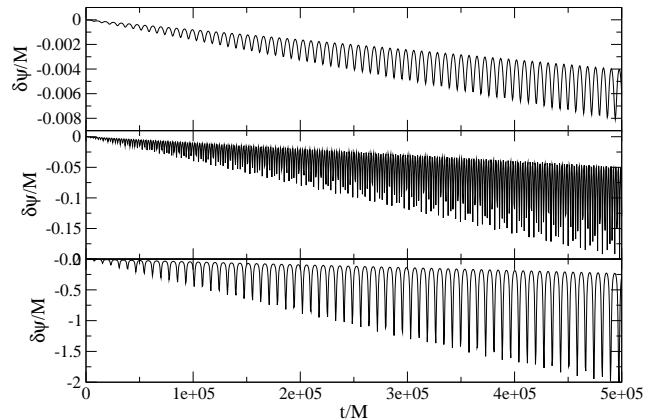}
\caption{\label{dephase-Psi} Dephasing of the $\psi$ coordinate as a function of time. The top panel corresponds to 
system $(A)$, the middle one to system $(B)$ and the bottom one to system $(C)$.}
\end{center}
\end{figure}

From this figure one cannot discern whether the CS trajectories trail or anticipate the GR ones, but this can be 
assessed by plotting the difference between the GR and CS evolved angles $(\psi,\chi,\phi)$ that determine 
completely the geodesic evolution. These angles are the Boyer-Lindquist azimuthal angle $\phi$ plus the two
angles $\psi$ and $\chi$ associated with the other Boyer-Lindquist coordinates $r$ and $\theta$ through the
relations in Eq.~\eqref{rzangles}. Figure~\ref{dephase-Psi} plots the dephasing in $\psi$, where we observe that, 
in all panels, the difference $\delta \psi := \psi_{\CS} - \psi_{\GR}$ is initially close to zero and then it 
decreases linearly on average (over a number of cycles). 
%%%%%%%%%%
%%%%%%%%%%
Also we observe that, as expected, system $(C)$ presents the biggest dephasing, followed by system $(B)$ and 
then by system $(A)$ (note the change of scale in the $y$-axis). Similar linear dephasing trends are observed 
in $\delta \chi$ and $\delta \phi$. 

For weakly gravitating systems, we find that $\delta \psi$ is dominant and negative, while 
$\delta \chi \sim -\delta \phi$, with the latter positive. This implies that the CS orbits of systems $(A)$ and 
$(B)$ are overall trailing the GR solution. This scenario is a bit more complicated for system $(C)$, where 
$\delta \psi \ll 0$ is still dominant, but  both $\delta \phi$ and $\delta \chi$ are positive, implying that 
the CS orbits still trail the GR one in radius, but anticipate it in angles $\theta$ and $\phi$.

%------------------------------------------------------------------------------------------------
\subsection{Gravitational Waveform}

\begin{figure}[htp]
\begin{center}
\includegraphics[scale=0.33,clip=true]{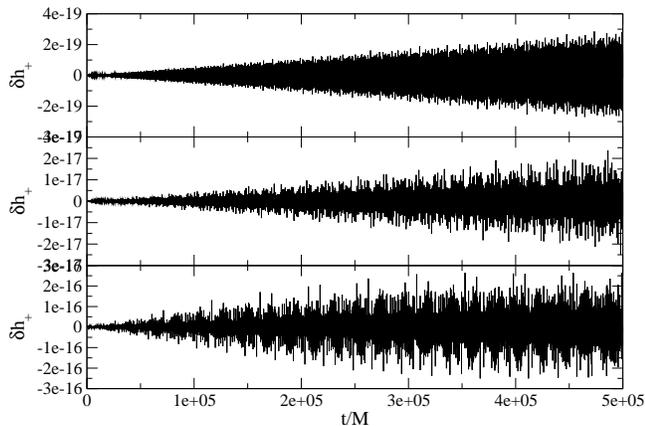}
\caption{\label{dephase-Hp} Dephasing of the GW polarization $h^{}_{+}$ as a function of time. 
The top panel corresponds to system $(A)$, the middle one to system $(B)$ and the bottom one to system 
$(C)$.}
\end{center}
\end{figure}

Orbital trajectories can give us a sense of the CS effect on geodesics, but the true observables are 
gravitational waveforms.  Figure~\ref{dephase-Hp} plots the difference in the waveforms computed 
with CS trajectories and GR trajectories.
For reference, the maximum magnitude of the GR GW polarization
$h^{}_{+}$ is approximately $|h^{}_{+}| < 8 \times 10^{-17}$ [system $(A)$], $|h^{}_{+}| < 1.25 
\times 10^{-16}$ [system $(B)$], and $|h^{}_{+}| <  2.25 \times 10^{-16}$ [system $(C)$].
%%%%%%%%%%
%%%%%%%%%%
At the end of the simulation ($\sim 128$ days of data using $M = M^{}_{\bullet}$), we observe that 
system $(A)$ has dephased by as much as $0.3 \%\,$, while system $(B)$ has dephased by $16 \%$, 
and system $(C)$ by $90 \%$ relative to the maximum amplitude of the respective GR waveforms. 
Similar behavior is observed for the other polarization. 

\begin{figure}[htp]
\begin{center}
\includegraphics[scale=0.33,clip=true]{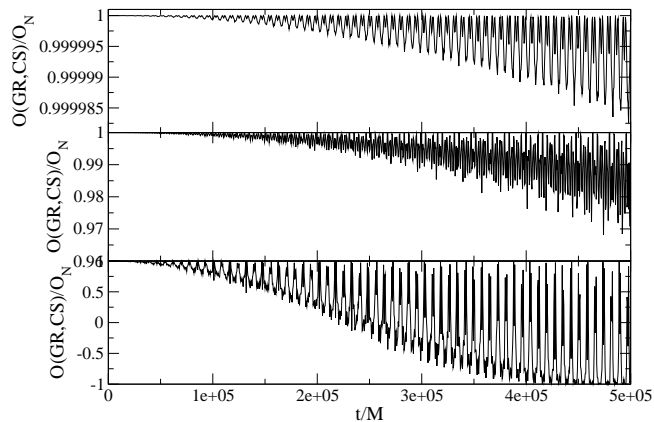}
\caption{\label{overl} Running average of the normalized dephasing measure $\bigcirc(\GR,\CS)$ over twenty 
consecutive time steps as a function of time. The top panel corresponds to system A, the middle one to system B 
and the bottom one to system C.}
\end{center}
\end{figure}

Another measure one can construct to observe this dephasing is the time-dependent overlap $\bigcirc(\GR,\CS)$, 
defined as
\be
\bigcirc(A,B) := h^{A}_{+} h^{B}_{+} + h^{A}_{\times} h^{B}_{\times}\,, \label{dephasingmeasure}
\ee
which we can normalize by using the quantity
\be
\bigcirc^{}_{N}(\GR,\CS) := \sqrt{\bigcirc(\GR,\GR)\,\bigcirc(\CS,\CS)}\,. \label{dephasingnormmeasure}
\ee
This measure is related to the integrand of the overlap commonly used in gravitational wave data analysis. 
Figure~\ref{overl} plots $\bigcirc(\GR,\CS)/\bigcirc^{}_{N}$ as a function of time.   
%%%%%%%%%%
%%%%%%%%%%
Observe that after only three weeks of data (roughly $10^{5} M$ of evolution), system $(A)$ and $(B)$ will 
probably not be able to distinguish between GR and CS geodesics. On the other hand, system $(C)$ will 
probably be able to distinguish this difference, because it experiences a deeper gravitational potential. 
This plot shows how robust LISA gravitational wave astronomy can be when considering geodesic motion about 
a spinning MBH, since even a rather strong curvature correction to the action leads to gravitational waves 
that are virtually indistinguishable from GR ones, unless the SCO is close to the light-ring or the coupling 
constant of the curvature correction is quite large.    

\begin{figure}[htp]
\begin{center}
\includegraphics[scale=0.33,clip=true]{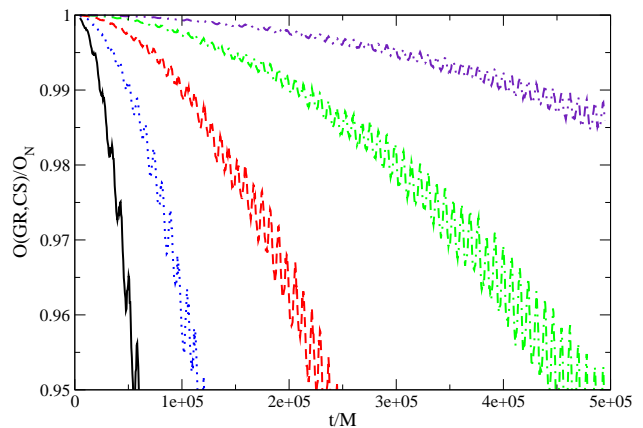}
\caption{\label{overl2} Normalized dephasing measure ${\cal{O}}(\GR,\CS)$ as a function of time for system $(C)$ 
using: 
$\xi = 0.4 M^{4}$ (solid black),
$\xi = 0.2 M^{4}$ (dotted blue),
$\xi = 0.1 M^{4}$ (dashed red), 
$\xi = 0.05 M^{4}$ (dot-dashed green),
and $\xi= 0.025 M^{4}$ (dot-dot-dashed violet).}
\end{center}
\end{figure}

By considering system $(C)$, the most relativistic one, one can ask how small can $\zeta$ be 
in order to cause a loss in the normalized dephasing measure of about $5\%\,$ after $10^{5}\,$M of evolution
[Eqs.~\eqref{dephasingmeasure} and~\eqref{dephasingnormmeasure}].
The answer can be found in Figure~\ref{overl2}, which shows the average of the normalized dephasing measure as a 
function of time for system $(C)$, but for the following values of the CS parameter $\zeta$: $\xi = 0.4 M^{4}$ (solid black), 
$\xi = 0.2 M^{4}$ (dotted blue), $\xi = 0.1 M^{4}$ (dashed red), $\xi = 0.05 M^{4}$ (dot-dashed green),
and $\xi = 0.025 M^{4}$ (dot-dot-dashed violet).  Observe that over 4 months of data, the CS correction could 
lead to a significant dephasing even for a CS parameter of ${\cal{O}}(10^{-2})$. On the other hand, if 
one integrates incoherently over 3 week segments of the data, then the CS correction would lead to a dephasing 
only if the CS parameter if of ${\cal{O}}(10^{-1})$. In this case, however, one would have trouble connecting 
three-week segments together if the CS correction is not taken into account.  

%%%%%%%%%%
%%%%%%%%%%

Another important question is how this bound compares with bounds obtained from binary pulsar tests. 
The precession of the perigee in binary pulsars can be 
used to argue that $\zeta^{1/4} \lesssim 10^{4} \; {\textrm{km}}$, as was done in Paper I.  
For inspirals, a GW test of DCSMG can be expressed as
\be
\xi^{1/4} \lesssim 6 \times 10^{6}\, \delta^{1/4}\, (M/M_{\bullet}) \, {\textrm{km}}\,,
\ee
where $\delta$ here represents the accuracy to which we can 
measure $\zeta$. This accuracy depends not only on the integration time, the signal-to-noise ratio and 
the type of orbit considered, but it should also be enhanced if one accounts for radiation-reaction effects, since 
they are also CS corrected (see Sec.~\ref{radiationreaction}). Based on the results of this section, it is 
not ludicrous to expect that $\delta \sim 10^{-6}$ or better depending on the system, while one also sees 
that IMRIs are favored over EMRIs due to the $M^{-1}$ dependence. Thus, if one considers an IMRI with total 
mass $M = 10^{3} M_{\odot}$, and if one assumes $\delta = 10^{-6}$, then the constraint can be better than 
the binary pulsar one by at least two orders of magnitude. Moreover, as we shall see in Sec.~\ref{radiationreaction}, 
radiation-reaction effects depend on the oscillatory sector of the CS scalar, which in turn affects GWs 
directly. Such an oscillatory sector of the theory is simply {\emph{untestable}} by binary pulsar experiments. 

The results and arguments presented above suggest that GW observations with LISA could allow for interesting 
tests of DCSMG, but of course, these arguments should be backed up by a deeper and more 
thorough data analysis study. In this sense, the figures presented in this section should only be taken as 
an indication of the possible dephasing and loss of overlap that one could experience if CS theory is in 
play. A detailed study of the dependence of such a test on the spectral noise density curve, the location 
of the detector in the sky via the beam pattern functions, and the distance to the source, is thus of key importance, but
postponed to future work.

%%%%%%%%%%%%%%%%%%%%%%%%%%%%%%%%%%%%%%
\section{Gravitational Wave Propagation and Radiation Reaction}\label{radiationreaction}

The semi-relativistic approximation provides a sensible approximation to the emission of GWs generated 
by the the motion of SCOs around MBHs, but so far radiation-reaction has been completed neglected.
These effects can be incorporated via the adiabatic 
approximation~\cite{Mino:2003yg,Sago:2005gd,Hughes:2005qb,Ganz:2007rf,Fujita:2009rr}, 
that is, the assumption that the changes in the {\em constants} of motion $(E,L,Q)$ evolve in a time
scale much larger than the orbital ones.  In this scheme, one then argues that the system can be evolved using
a geodesic evolution for a certain number of orbital cycles, after which the constants of motion are corrected
by estimates that use balance laws (eg.~the GR energy balance law provides a prescription of how to modify 
the orbital energy due to GW energy emission out to infinity and into the horizon).

In order to obtain the balance laws one must understand the propagation of GWs in the underlaying theory and
their associated, effective stress-energy tensor.  This can be done in the framework of the
shortwave approximation, where one decomposes the geometry into a background and an oscillatory
part corresponding to GWs. Such a decomposition holds when the GW wavelength is much smaller than the
typical length scale associated with the background curvature.

In this section, we develop the shortwave approximation for DCSMG and compute the effective
stress-energy tensor of GWs in this theory.  Based on this, and using the symmetries of
the background geometry (timelike and axial Killing vectors), we can establish balance laws that
can allow us to introduce radiation-reaction effects, in the adiabatic approximation, in the
energy and angular momentum.  For the case of the Carter constant the situation is more difficult
as it is not associated with a Killing vector symmetry, but perhaps a two-timescale expansion could
be performed, similar to that introduced in~\cite{Flanagan:2007tv}.

%---------------------------------------------------------------------------------------------------------------
\subsection{The Short-Wavelength Approximation}

The short-wavelength approximation (SWA) was first introduced by Isaacson~\cite{Isaacson:1968ra,Isaacson:1968gw} 
(see also~\cite{Misner:1973cw} and references therein) to study the propagation of GWs beyond the linear approximation.   
In the SWA one assumes that the typical GW wavelength, $\lgw$,  is much smaller than the curvature length-scale, 
${\cal R}$, such that $\lgw \ll {\cal R}$, where ${\cal R}^{-2}$ is the typical scale 
of the components of the Riemann tensor.  In this way, we can split the spacetime metric into a background metric,
$\bar{\met}^{}_{\mu\nu}$, with associated curvature scale ${\cal R}$, 
plus an {\em oscillating} metric perturbation describing the GWs, $h^{}_{\mu\nu}$, with an associated length 
scale $\lgw$ [see eg.~Eq.~\eqref{splitting}]
and an associated amplitude $\hgw$.   In this way, $\partial_{\alpha}\bar\met_{\mu\nu}\sim 1/{\cal R}$ and  
$\partial_{\alpha} h_{\mu\nu} \sim \hgw/\lgw$. 

In practice, applying the SWA to any tensor implies separating its background part from its oscillatory part, 
which can be done by averaging over several wavelengths ($\lgw$).  Such separation can be accomplished via 
the Brill-Hartle averaging procedure~\cite{Isaacson:1968ra,Isaacson:1968gw},
which applies to convex regions of the background spacetime (regions where any two points can be joined by a 
unique geodesic).  Then, the average of the components of a given tensor $T^{a_{1}\cdots}{}_{b_{1}\cdots}$ is defined as:
\begin{widetext}
\be
<T^{a_{1}\cdots}_{b_{1}\cdots}>(x) := \int^{}_{{\cal W}} d^{4}x'\,\sqrt{-\bar{\met}(x')}\; f(x',x)\,
\hat\met^{a_{1}}_{a'_{1}}(x,x')\cdots
\hat\met_{b_{1}}^{b'_{1}}(x,x')\cdots T^{a'_{1}\cdots}_{b'_{1}\cdots}(x')  \,,
\ee
\end{widetext}
where the integration takes place over a {\em small} region, ${\cal W}$, containing several GW wavelengths;
$\hat\met^{a_{1}}{}_{a'_{1}}$ is the bitensor of parallel displacement in the background spacetime;
and $f(x',x)$ is a weighting function that smoothly decays to zero as $x$ and  $x'$ are separated by
many wavelengths.   The main properties of this averaging rule (for
details see~\cite{Misner:1973cw}) are:
\begin{enumerate}
\item Background covariant derivatives commute [fractional errors are of ${\cal{O}}(\lgw/{\cal R})^2$].
\item Total divergences average out to zero [fractional errors are of ${\cal{O}}(\lgw/{\cal R})$].
\item As a corollary of (1) and (2), one can freely integrate by parts.
\end{enumerate}
These rules shall be used heavily in the next section to obtain an expressions for the GW stress-energy tensor
in CS modified gravity.

%---------------------------------------------------------------------------------------------------------------
\subsection{Effective Stress-Energy Tensor for GWs}\label{Eflux}

In DCSMG, not only does the metric tensor oscillate, but also the CS scalar field, sourced exclusively 
by the spacetime curvature [see Eq.~\eqref{EOM}].  When GWs are 
present, these will then generically induce oscillations in the CS scalar field, which forces us to
also decompose $\vartheta$ into a background part, $\bar\vartheta$, 
and an oscillatory part, $\tilde\vartheta$, the latter induced by GWs.

Let us now study the field equations in the SWA, which requires expansions
up to second order in the oscillatory parts (see Appendix~\ref{2ndswa} for the basic formulae used for these
expansions).  The structure of the resulting equations consists of 
oscillatory terms, linear in the amplitude, and {\em coarse-grained} terms that describe how the 
background is modified by GWs.  This modification is produced by the {\em averaged} second-order
term in the oscillatory part, which reflects the nonlinear character of this effect.   
 
Then we set to zero the linear part in the wave amplitude, $\hgw$, obtaining in this way the equations that
describe the propagation of the waves in the background. 
In GR, these equations are simply
\ba
\oone R^{}_{\mu\nu} & = & -\frac{1}{2}h_{\mu \nu|\rho}{}^{\rho} 
+ {\cal O}\left(\frac{\hgw}{{\cal R}^{2}}\right)  \\ \nonumber
& = &\frac{1}{2\kappa}\left( \oone T^{\MAT}_{\mu\nu}
-\frac{1}{2}h^{}_{\mu\nu}T^{\MAT} - \frac{1}{2}\bar\met^{}_{\mu\nu} \oone T^{\MAT} \right) \,,
\ea
where we have used the Lorenz gauge~[Eq.\eqref{lorenzgauge}] and 
where, here and in the rest of this section, the superscript preceding a given quantity denotes the
perturbative order of this quantity with respect to metric perturbations (see Appendix~\ref{appswa} for 
more details).  

In DCSMG the propagation equations have the following form
\ba
& &\oone R^{}_{\mu\nu} + \frac{\alpha}{\kappa}\left( \oone C^{}_{\mu\nu}[\bar\vartheta\,,h]
+ C^{}_{\mu\nu}[\tilde\vartheta\,,\bar\met]\right) = \frac{1}{2\kappa}\left[ \oone T^{\MAT}_{\mu\nu} 
\right.\nonumber \\ 
& & \left. + \oone T^{(\vartheta)}_{\mu\nu}[\bar\vartheta]
+ \delta T^{(\vartheta)}_{\mu\nu}[\bar\vartheta,\tilde\vartheta]
-\frac{1}{2}h^{}_{\mu\nu}(T^{\MAT}+T^{(\vartheta)}[\bar\vartheta])\right. \nonumber \\
&& \left. - \frac{1}{2}\bar\met^{}_{\mu\nu}( \oone T^{\MAT} 
+ \oone T^{(\vartheta)}[\bar\vartheta] + \delta T^{(\vartheta)}[\bar\vartheta,\tilde\vartheta])\right]\,, 
\label{swagwprop}
\ea
\ba
\beta\left\{ \bar\square \tilde\vartheta - \bar\met^{\alpha\beta}\oone S^{\lambda}_{\alpha\beta}
\bar\vartheta^{}_{|\lambda} - h^{\alpha\beta}\bar\vartheta^{}_{|\alpha\beta} \right\} \nn \\
=  -  \frac{\alpha}{2}\left\{ {\,^\ast\!}\bar{R}\; \oone\!R[h] + \frac{1}{4}h{\,^\ast\!}\bar{R}\;
\bar{R}\right\}\,. \label{swacsprop}
\ea
The detailed structure of Eq.~\eqref{swagwprop} as well the form of 
$\delta T^{(\vartheta)}_{\mu\nu}[\bar\vartheta,\tilde\vartheta]$ can be found in Appendix~\ref{linearswa} 
[Eq.~\eqref{firstorderequations}].  The object $\oone S^{\mu}_{\alpha\beta}$ in Eq.~\eqref{swacsprop} 
denotes the perturbation of the Christoffel symbols and its expression is given in Eq.~\eqref{onestensor}. 

Using the averaging procedure described above, the field equations tell us how the nonlinear contribution
of the waves shapes the background.  This contribution can be written as an effective stress-energy
tensor of the GWs, $T^{\GW}_{\mu\nu}$ (the Isaacson tensor), allowing us to write
\be 
\bar{G}^{}_{\mu\nu} + \frac{\alpha}{\kappa}\bar{C}^{}_{\mu\nu} = \frac{1}{2\kappa}\left( T^{\MAT}_{\mu\nu}
+ T^{(\vartheta)}_{\mu\nu}[\bar\vartheta]+ T^{(\vartheta)}_{\mu\nu}[\tilde\vartheta] + T^{\GW}_{\mu\nu}\right)\,.
\label{csswaback1}
\ee
In GR, the Isaacson tensor is given by~(see e.g.~\cite{Misner:1973cw})
\be
T^{\GW}_{\mu\nu} = -2\kappa\left\{ < \otwo R^{}_{\mu\nu}[h]>
-\frac{1}{2}\bar{\met}^{}_{\mu\nu}<\otwo R[h]>\right\}\,, \label{tgw}
\ee
which, using the expressions in Appendix~\ref{appswa} and working in the Lorenz gauge
[Eq.~\eqref{lorenzgauge}] supplemented
with the traceless condition [Eq.~\eqref{tracelesscondition}], 
reduces to:
\be
T^{\GW}_{\mu\nu} = \frac{\kappa}{2}\left<  {h}_{\alpha \beta | \mu} {h}^{\alpha \beta}{}_{|\nu}\right>\,.
\label{GRTab}
\ee
One can see, using the first-order equations and the average rule, that this tensor is traceless and divergence-free
within the errors produced by the approximations made using the SWA scheme:
\be
\bar\met^{\mu\nu}T^{\GW}_{\mu\nu} = 0\,, \qquad
T^{\GW}_{\mu\nu}{}^{|\nu} = 0\,.
\ee

In DCSMG, the effective stress-energy tensor of GWs is given by
\ba
T^{\GW}_{\mu\nu} & = & -2\kappa\left\{ < \otwo R^{}_{\mu\nu}[h]>
-\frac{1}{2}\bar{\met}^{}_{\mu\nu}< \otwo R[h]> \right. \nn \\
& + & \left. \!\!\! \frac{\alpha}{\kappa}\left( <\otwo C^{}_{\mu\nu}[\bar\vartheta\,,h]> + 
<\oone C^{}_{\mu\nu}[\tilde\vartheta\,,h]>\right) \right\}\,.
\label{cstgw}
\ea
Again, using the formulae in Appendix~\ref{appswa}, restricted to the TT gauge of Eqs.~\eqref{lorenzgauge}
and~\eqref{tracelesscondition}, and the properties of the averaging (in particular the integration by parts),
we find that the terms corresponding to the CS correction {\emph{vanish exactly}}, yielding  the same expression as in GR
(for a guided explanation of this fact, see Appendix~\ref{appswa}).  
That is, the effective stress-energy tensor of GWs in DCSMG (which enters in Eq.~\eqref{csswaback1}) 
is simply given by Eq.~\eqref{tgw} or~\eqref{GRTab}, which implies that the backreaction in the orbital motion due to GW emission
is essentially as in GR. 

We can now look at the equation for $\vartheta$ with the second-order corrections.  This is a linear
equation in $\vartheta$ and hence all the modifications come from second-order terms in the GW.  
One the finds that
\ba
\beta\bar\square\bar\vartheta & = & - \frac{\alpha}{4}  \left(1 - 
\frac{1}{4}\left< h^{\alpha\beta}h^{}_{\alpha\beta}\right>\right) {\,^\ast\!} \bar{R}\;\bar{R}  \nn \\
& - &  \frac{\alpha}{4} \bar\epsilon_{\alpha\beta\mu\nu}\left< \oone R^{\alpha\beta\rho\sigma}\; 
\oone R_{\rho\sigma}{}^{\mu\nu}\right> \,. \label{csswaback2}
\ea
As we can see there are two second-order corrections to the background value of the 
CS scalar field: one that {\em renormalizes} the value of $\alpha$, $<\!\! h^{\alpha\beta}h^{}_{\alpha\beta}\!\!>\;
\sim h^{2}_{\GW}$, and another that scales as $\alpha\, h^{2}_{\GW}/\lambda^{4}_{\GW}$.  

%---------------------------------------------------------------------------------------------------------------
\subsection{Balance Laws and Radiation-Reaction Effects}\label{RR}

By means of the SWA we have studied the propagation of GWs and of oscillations induced in the CS 
scalar field.  We have also seen how these oscillations affect the background values of the metric
tensor [Eq.~\eqref{csswaback1}] and of the CS scalar field [Eq.~\eqref{csswaback2}].   
Of particular importance is the equation for the modification of the background geometry, which
contains the effective stress-energy tensor of the GWs, which we have found has the same form as in
GR.  From the previous development it is clear that from Eq.~\eqref{csswaback1} we have the following
conservation equation
\be
\bar\nabla^{\nu}\tau^{}_{\mu\nu} = 0\,,~~
\tau^{}_{\mu\nu} = T^{\MAT}_{\mu\nu} + T^{(\vartheta)}_{\mu\nu}[\tilde\vartheta] 
+ T^{\GW}_{\mu\nu} \,,  \label{effconservation}
\ee
where we have used that the background Einstein tensor is divergence-free and the equations of 
motion for the background scalar field $\bar\vartheta$ are satisfied.  The new ingredient with respect to GR is
the appearance of the stress-energy tensor associated with the oscillations induced in the CS
scalar field.

Here,  the background clearly corresponds to the MBH geometry and
CS background scalar field is given in Eqs.~\eqref{sol:metric_elements} and~\eqref{back-theta}.  Using the 
Killing symmetries of the background, described by the vector fields $t^{\mu}$ and $\psi^{\mu}$ 
[see Sec.~\ref{geodesics}], the vector fields
\be
{\cal E}^{\mu} = -\tau^{\mu}{}^{}_{\nu}t^{\nu}\,, \qquad
{\cal J}^{\mu} = \tau^{\mu}{}^{}_{\nu}\psi^{\nu}\,, \label{vectorfluxes}
\ee
describe the total fluxes of energy and angular momentum and are, by virtue of Eq.~\eqref{effconservation}
and the Killing equations, divergence-free vector fields.   Therefore, we can obtain balance laws
for the energy and angular momentum by considering a spacetime region ${\cal V}$ with boundary 
$\partial{\cal V}$ and integrating over it the divergence-free conditions: ${\cal E}^{\mu}{}_{|\mu} = 0$
and ${\cal J}^{\mu}{}_{|\mu} = 0$.   

For the study of IMRI/EMRIs one can take ${\cal V} = 
\{ (t,r,\theta,\phi)\,,~\mbox{such that}~t^{}_{o}<t<t^{}_{f}~\mbox{and}~r^{}_{H}<r<r^{}_{I}\}$,
where the inner radius, $r^{}_{H}$, can be taken to be close to the horizon location and the outer
radius, $r^{}_{I}$, can be taken to be effectively close to infinity (or at least far away from 
the SCO).  In this way $\partial{\cal V}$ is composed of two {\em cylinders} (one at $r=r^{}_{H}$
and the other one at $r=r^{}_{I}$) of finite size, limited by two slices, one at an initial time 
$t=t^{}_{o}$ and the other one at a final time $t=t^{}_{f}$.  

The balance laws for energy and angular momentum tell us that the change in the constants
of motion $(E,L)$ is given by the amount energy/angular momentum, of the GWs and of the CS scalar field, 
flowing away from ${\cal V}$ to infinity ($r=r^{}_{I}$) and into the MBH horizon ($r=r^{}_{H}$).
In this way, one can account for the radiation-reaction effects in the adiabatic
approximation.  Note that this only takes care of the changes in the energy and in the angular momentum
of the SCO, while the question of how to correct the third constant of motion,
the Carter constant $Q$, in a general way remains still open
(see~\cite{Sago:2005gd,Hughes:2005qb,Drasco:2005kz,Ganz:2007rf,Fujita:2009rr} for recent advances on that question). 

The flux of energy (or energy luminosity) in GWs going towards infinity is given by
\be
\frac{dE^{}_{\GW}}{dt} = - \lim^{}_{r\rightarrow\infty} r^{2}\int^{}_{S_{\infty}^{2}} \!\!\!\!
d\Omega \;T^{\GW}_{t i}\,n^{i} \,, \label{eluminosity}
\ee
where $n^{i} = x^{i}/r$ is the unit normal to the surfaces $r={\textrm{const.}}$ near spatial infinity,
where the background geometry is essentially flat.  Although the expression for the energy
luminosity [Eq.~\eqref{eluminosity}] is formally the same as in GR, the luminosities are actually different 
because the geodesic equations are indeed CS corrected by the modified Kerr background
given in Eq.~\eqref{sol:metric_elements}.

\begin{figure}[htp]
\begin{center}
\includegraphics[scale=0.33,clip=true]{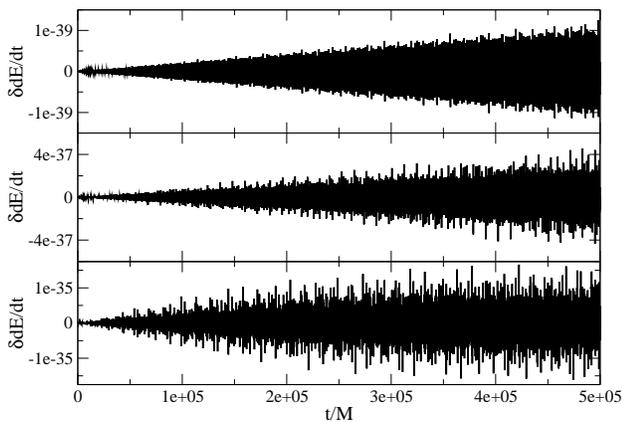}
\caption{\label{luminos} Energy flux difference (in units of $M^{-2}$) between the flux computed with 
gravitational waves produced by geodesics about Kerr and about the CS modified Kerr background for 
systems $(A)$ (top panel), $(B)$ (middle panel) and $(C)$ (bottom panel) as a function of time.}
\end{center}
\end{figure}

To illustrate this difference, let us consider the expression
\be
\frac{dE^{}_{\GW}}{dt} = \kappa\left( \dot{h}^{2}_{+} + \dot{h}^{2}_{\times}\right)\,,
\ee
where $h_{+}$ and $h_{\times}$ are the GW plus- and cross-polarizations, which one can 
obtain from Eqs.~\eqref{GRTab} and~\eqref{eluminosity}. 
Figure~\ref{luminos} plots the difference between the energy flux computed 
from the GW emission of a SCO moving in geodesics of Kerr and a SCO moving in geodesics of 
the CS-modified Kerr metric [Eq.~\eqref{sol:metric_elements}], for systems $(A)$ (top panel), $(B)$ 
(middle panel) and $(C)$ (bottom panel) as a function of time.  
For reference, the maximum magnitude of the energy flux computed from geodesic waveforms about Kerr are 
$3 \times 10^{-37} M^{-2}$ for system $(A)$, $3 \times 10^{-36} M^{-2}$ for system $(B)$ and 
$2 \times 10^{-35} M^{-2}$ for system $(C)$.
%%%%%%%%%%
%%%%%%%%%%
Observe that, after approximately four months of evolution, the fractional correction to the luminosities 
by the CS correction is about $1\%$ for system $(A)$, $10\%$ for system $(B)$ and about $50\%$ for system $(C)$. 
As expected, the more relativistic the system, the larger the CS correction to the waveforms, and therefore, the larger 
the effect on radiation-reaction. Such corrections could have a significant impact on kludge-type waveforms, when 
one takes into account that the GW energy lost must also be supplemented by the energy emitted
by the CS scalar field [see the conservation equation, Eq.~\eqref{effconservation}].  To include
this contribution, one would need to solve the evolution equation for the CS scalar field [Eq.~\eqref{1stfieldpert}], 
using as input the gravitational waveform computed without radiation-reaction. 

The machinery developed in this paper allows, for the first time, the inclusion of these radiation-reaction 
effects in the orbital evolution of EMRIs and IMRIs in a realistic, alternative theory of gravity.
With these developments, one can now properly construct alternative-theory, kludge waveforms.

%%%%%%%%%%%%%%%%%%%%%%%%%%%%%%%%%%%%%%
\section{Conclusions}
\label{conclusions}

We have studied the effects of dynamical CS modified gravity on the orbit and GWs generated by IMRIs and EMRIs. 
The semi-relativistic approximation was employed, through which we modeled the SCO trajectory via the geodesic
equations, while we treated GWs through a multipolar generation formalism. 

We began by showing that test particles follow geodesics in DCSMG. We then proceeded to 
explicitly calculate the modified geodesic equations for a test particle in a generic orbit around a CS 
modified Kerr MBH~\cite{Yunes:2009hc}.  We showed that the fundamental frequencies of the modified background 
are CS modified, as well as the relations between the MBH multipole moments and its mass and spin, 
although the latter incurs corrections at $\ell=4$ multipolar order, which might be small to be 
detected by future GW observations with LISA. 

We then continued with the study of GW generation in the modified theory. We saw that the GW generation
formula based on the multipolar expansion is corrected to leading-order by second-order terms of ${\cal{O}}(\mu \zeta)$, 
which we neglect here.  These GW generation formalism is also modified by GR radiation-reaction effects, which 
scale as the square of the mass ratio and which we also neglect here.  

We then evolved geodesics in this modified background and compared their associated GWs to those generated 
by geodesics on a pure Kerr background. Generically, we find that these waves dephase after a certain number of 
cycles that depends on the strength of the CS coupling parameter, as well as on how relativistic the geodesic under
consideration is and the signal to noise ratio of the event. 
For small CS parameters, such a dephasing is small, leading to mismatches that will not affect 
GW detection. These mismatches, however, might affect GW characterization, leading to erroneous conclusions
about the orbital parameters of the system.    

EMRIs and IMRIs can thus be used to probe DCSMG, both by constraining the CS coupling 
strength to unprecedented levels, as well as to sample the oscillatory dynamic sector of the CS scalar 
field. Moreover, the study presented can be viewed as an explicit example of the effect a non-Kerr 
background could have on GWs, where in this case the non-Kerrness arises from a well-defined and 
physically well-motivated alternative theory.  Since this modified theory introduces a higher-order 
curvature correction, modifications to the Kerr metric are only dominant in the strong-field and cannot 
be sampled by the waveform's first multipoles.   

The analysis presented here employs some approximations that leads to certain inaccuracies. These
inaccuracies are associated with: (i) neglecting radiation-reaction in the dynamics, which
scale with the square of the mass ratio; (ii) neglecting CS modifications to the GW emission, which scale with the 
product of the mass ratio and the CS coupling constants; (iii) neglecting CS modifications to the MBH background metric
that are of third-order in the spin parameter and the small-coupling approximation.
Whether one effect is dominant over the others depends on the strength of the CS coupling constants, 
which is at present unknown.

Future work could concentrate on the inclusion of radiation-reaction in an adiabatic fashion to the 
waveforms presented here. Such a task in the modified theory is rather straightforward since these 
dissipative effects are CS corrected only through the emission of CS scalar field. 
One could then supplement the standard GR 
expressions for the rate of change of the geodesic constants of motion by these CS corrections
to allow for a true, kludge-like, inspiral. 
Although the CS correction does not modify radiation-reaction through changes in the Isaacson tensor, 
dissipative effects will exentuate the observed dephasing.

Once such radiation-reaction effects are included, a thorough data analysis study should be carried out 
to confirm or not the expectations presented in this paper. Of particular importance will be the inclusion of 
the beam pattern functions and the motion of the detector in the sky, which have been shown to be relevant for 
parameter estimation~\cite{Will:2004xi,Berti:2004bd,Berti:2005qd,Arun:2009pq}. Moreover, the computation of the 
proper overlap and Fisher matrix would include the spectral noise density curve of the detector, as well as 
details of the sensitivity of the modified waveforms on the parameters of the alternative theory. 

Another avenue for future research is the inclusion of first, leading-order CS correction to the GW emission 
framework. We have provided here a map, albeit involved, to construct such a correction which is based 
on black hole perturbation theory. Another possibility would be to study GW generation in the context of a 
post-Newtonian/post-Minkowskian expansion and asymptotic 
matching~\cite{Blanchet:1989ki,Damour:1990ji,Blanchet:1995fr,Blanchet:1998in,Poujade:2001ie}. Although this formalism is formally 
valid in the slow-motion approximation, so is the truncated multipolar decomposition used here. 

Last but not least, an exact metric that describes spinning MBHs with arbitrary spin angular momentum in 
DCSMG is still lacking. The results presented here are in principle valid only in the slow-rotation approximation. 
Generalizing such results to higher order in a perturbative fashion could be dramatically difficult, since to 
next order the scalar stress-energy tensor will also contribute to the modified field equations. Perhaps, a full 
numerical study of the collapse of a scalar field with angular momentum would be appropriate to disentangle the 
effects of DCSMG on rapidly rotating MBH backgrounds.

%%%%%%%%%%%%%%%%%%%%%%%%%%%%%%%%%%%%%%%%%%%%%%%%%%%%%%%%%%%%%%%%%%%%%%%%%%%%%%
\acknowledgments We would like to thank Bernard Schutz
for encouraging us to study EMRIs in the context of 
CS modified gravity.  We would also like to thank Stephon
Alexander, Emanuele Berti, Vitor Cardoso, Daniel Grumiller, Scott Hughes, 
Ben Owen, Frans Pretorius, David Spergel, Paul Steinhardt and Shaun Wood, 
for helpful discussions. 

CFS acknowledges support from the Ram\'on y Cajal Programme of the
Ministry of Education and Science of Spain, by a Marie Curie
International Reintegration Grant (MIRG-CT-2007-205005/PHY) within the
7th European Community Framework Programme, and from the contract ESP2007-61712
of the Spanish Ministry of Education and Science. He also acknowledges the
hospitality of the Physics Department at Princeton University during a visit
where this work was completed, and the CSIC for financial suport for this visit.  
NY acknowledges support from NSF grant PHY-0745779 and also acknowledges
the hospitality of the Institut de Ci\`encies de l'Espai (CSIC-IEEC) during
the realization of this work. 

%%%%%%%%%%%%%%%%%%%%%%%%%%%%%%%%%%%%%%%%%%%%%%%%%%%%%%%%%%%%%%%%%%%%%%%%%%%%%%

\appendix

%%%%%%%%%%%%%%%%%%%%%%%%%%%%%%%%%%%%%%%%%%%%%%%%%%%%%%%%%%%%%%%%%%%%%%%%%%%%%%
\section{Formulae for the short wavelength approximation} \label{appswa}
In this Appendix we provide some of the formulae that we have computed and 
used to derive the form of the effective stress-energy tensor of  
GWs for DCSMG, using the SWA.  However,
these are generic perturbative expansions (induced only by perturbations
of the metric tensors) of geometric quantities which can be used in other 
perturbative schemes, although the interpretation of the different terms would be different.

%---------------------------------------------------------------------------------------------------------------
\subsection{Second-order expansion}\label{2ndswa}
Given the splitting of the spacetime metric given in Eq.~\eqref{splitting},
we need to perform an expansion in $h^{}_{\mu\nu}$ to second order 
of the different geometric quantities involved in the CS modified gravitational field
equations.  The inverse of the metric is:
\be
\met^{\mu\nu} = \bar{\met}^{\mu\nu} - h^{\mu\nu} + h^{\mu}{}_{\rho}
h^{\rho\nu} + {\cal O}(h^3)\,,
\ee
where
\be
h^\mu{}_\nu = \bar{\met}^{\mu\rho} h^{}_{\rho\nu}\,, ~
h_\mu{}^{\nu} = \bar{\met}^{\nu\rho} h^{}_{\mu\rho}\,, ~
h^{\mu\nu} = \bar{\met}^{\mu\rho} \bar{\met}^{\nu\sigma} h^{}_{\rho\sigma}\,.
\ee
The expansion of the metric determinant, $\met = \det(\met^{}_{\mu\nu})$,
is then given by:
\be
\met = \bar{\met} \left\{ 1 + h + \frac{1}{2}\left( h^2 - h^{\mu\nu}h_{\mu\nu}  \right)
+ {\cal O}(h^3) \right\}\,.    
\ee
From here we can expand the completely antisymmetric Levi-Civita volume
four-form:
\be
\epsilon^{}_{\mu\nu\rho\sigma} = \sqrt{-\met}\,\delta^{0123}_{\mu\nu\rho\sigma}\,,
\ee
to get
\be
\epsilon^{}_{\mu\nu\rho\sigma} = \bar\epsilon^{}_{\mu\nu\rho\sigma}\left\{
1 + \frac{1}{2}h + \frac{1}{8}(h^2-2h^{\alpha\beta}h^{}_{\alpha\beta}) 
+ {\cal O}(h^3) \right\}\,. 
\ee
The Christoffel symbols can be written exactly as follows:
\be
\Gamma^\mu_{\rho\sigma} = \bar\Gamma^\mu_{\rho\sigma} +
S^{\mu}_{\rho\sigma}\,,
\ee
where $S^{\mu}_{\rho \sigma}$ is
\be
S^{\mu}_{\rho\sigma} =  \frac{1}{2}\met^{\mu\nu}
\left( h^{}_{\sigma\nu |\rho} + 
h^{}_{\rho\nu|\sigma}-
h^{}_{\rho\sigma|\nu} \right)\,,
\ee
This tensor can be expanded in $h^{}_{\mu\nu}$ yielding
\be
S^\mu_{\rho\sigma} = \oone S^{\mu}_{\rho\sigma} + 
\otwo S^{\mu}_{\rho\sigma} + {\cal O}(h^3)\,,
\ee
where
\ba
\oone S^{\mu}_{\rho\sigma} &=&  \frac{1}{2}\bar{\met}^{\mu\nu}
\left( h^{}_{\sigma\nu |\rho} + 
h^{}_{\rho\nu|\sigma}-
h^{}_{\rho\sigma|\nu} \right)\,, \label{onestensor}
 \\
\otwo S^{\mu}_{\rho\sigma} &=&  -\frac{1}{2}h^{\mu\nu}
\left(
h^{}_{\sigma\nu |\rho} + 
h^{}_{\rho\nu|\sigma}-
h^{}_{\rho\sigma|\nu} \right)\,.
\ea
The introduction of the tensor $S^\mu_{\rho\sigma}$
is convenient since it can be used to simplify the 
expressions for the expansions of the Riemann tensor and
derived geometric quantities.  An exact expression relating
the Riemann tensor of the spacetime to the background one
through the tensor $S^\mu_{\rho\sigma}$ is:
%
%\begin{widetext}
\be
R^\alpha{}^{}_{\beta\mu\nu} = \bar{R}^\alpha_{}{}^{}_{\beta\mu\nu}
- 2\, S^{\alpha}_{\beta[\mu|\nu]}
+ 2\, S^{\alpha}_{\rho[\mu}S^{\rho}_{\nu]\beta}\,.
\ee
%\end{widetext}
%
Therefore, an exact expression for the Ricci tensor is:
%
%\begin{widetext}
\be
R^{}_{\beta\nu} = \bar{R}^{}_{\beta\nu}
+ S^{\alpha}_{\beta\nu|\alpha}
- S^{\alpha}_{\alpha\beta|\nu}
+ S^{\alpha}_{\alpha\rho}S^{\rho}_{\beta\nu}
- S^{\alpha}_{\rho\nu}S^{\rho}_{\beta\alpha}\,.
\ee
%\end{widetext}
%
From these expressions of the Riemann and Ricci tensors we can immediately
write down their first- and second-order terms in the perturbative expansion
in $h^{}_{\mu\nu}$.  For the Riemann tensor they are:
\be
\oone R^\alpha{}^{}_{\beta\mu\nu} = 
-2 \,\oone S^{\alpha}_{\beta[\mu|\nu]}\,,
\ee
\be
\otwo  R^\alpha{}^{}_{\beta\mu\nu} = 
-2 \,\otwo S^{\alpha}_{\beta[\mu|\nu]}
+ 2\, \oone S^{\alpha}_{\rho[\mu}\,\oone S^{\rho}_{\nu]\beta}\,,
\ee
and for the Ricci tensor:
\be
\oone R^{}_{\beta\nu} = 
\oone S^{\alpha}_{\beta\nu|\alpha}
- \oone S^{\alpha}_{\alpha\beta|\nu} \,,
\ee
\be
\otwo R^{}_{\beta\nu} = 
\otwo S^{\alpha}_{\beta\nu|\alpha}
- \otwo S^{\alpha}_{\alpha\beta|\nu}
+ \oone S^{\alpha}_{\alpha\rho}\,\oone S^{\rho}_{\beta\nu}
- \oone S^{\alpha}_{\rho\nu}\,\oone S^{\rho}_{\beta\alpha}\,.
\ee
The Einstein and Cotton tensors have more complicated expressions.
The first- and second-order expansion of the Einstein tensor are:
\be
\oone G^{}_{\mu\nu} = \oone R^{}_{\mu\nu} - \frac{1}{2}\bar{\met}^{}_{\mu\nu}
\left( \bar{\met}^{\rho\sigma}\,\oone R^{}_{\rho\sigma} - h^{\rho\sigma}
\bar R^{}_{\rho\sigma} \right) - \frac{1}{2}h^{}_{\mu\nu}\bar R \,,
\ee
\begin{widetext}
\be
\otwo G^{}_{\mu\nu} = \otwo R^{}_{\mu\nu} - \frac{1}{2}\bar{\met}^{}_{\mu\nu}
\left( \bar{\met}^{\rho\sigma}\,\otwo R^{}_{\rho\sigma} - h^{\rho\sigma}
\oone R^{}_{\rho\sigma} + h^{\rho}{}_{\lambda}h^{\lambda\sigma}\bar R^{}_{\rho\sigma}
\right) - \frac{1}{2}h^{}_{\mu\nu}\left(\bar{\met}^{\rho\sigma} \oone R^{}_{\rho\sigma}
-h^{\rho\sigma} \bar R^{}_{\rho\sigma} \right)\,.
\ee
\end{widetext}
The expansion of the first piece of the Cotton tensor, $C^{1}_{\mu\nu}$, is determined 
by the following expressions [we remark that these are expansions only in the metric
tensor, the CS scalar field is assumed here to be fixed]:
\begin{widetext}
\be
\oone C^{1}_{\mu\nu} = \vartheta^{}_{|\sigma}\bar{\epsilon}^{}_{(\mu}{}^{\sigma\alpha\beta}
\left\{ \oone R^{}_{\nu)\beta|\alpha} - \oone S^{\rho}_{\nu)\alpha}\bar R^{}_{\beta\rho} 
\right\} +  \vartheta^{}_{|\,\sigma} \oone{\epsilon}^{}_{(\mu}{}^{\sigma\alpha\beta}
\bar{R}^{}_{\nu)\beta|\alpha} \,, \label{1stC1}
\ee
\ba
\otwo C^{1}_{\mu\nu} & = & \vartheta^{}_{|\sigma}\bar{\epsilon}^{}_{(\mu}{}^{\sigma\alpha\beta}
\left\{ \otwo R^{}_{\nu)\beta|\alpha} - \oone S^{\rho}_{\nu)\alpha}\oone R^{}_{\beta\rho}
-\otwo S^{\rho}_{\nu)\alpha} \bar R^{}_{\beta\rho} 
\right\} +  \vartheta^{}_{|\sigma} \oone{\epsilon}^{}_{(\mu}{}^{\sigma\alpha\beta}
\left\{ \oone{R}^{}_{\nu)\beta|\alpha} - \oone S^{\rho}_{\nu)\alpha}\bar R^{}_{\beta\rho} 
\right\} \nonumber \\
& + & \vartheta^{}_{|\sigma} \otwo{\epsilon}^{}_{(\mu}{}^{\sigma\alpha\beta}
\bar{R}^{}_{\nu)\beta|\alpha} \,,
\ea
\end{widetext}
where $\oone{\epsilon}^{\sigma}{}_{\mu}{}^{\alpha\beta}$ and $\otwo{\epsilon}^{\,\sigma}{}_{\mu}{}^{\alpha\beta}$ 
are the first- and second-order expansion terms of ${\epsilon}^{\,\sigma}{}_{\mu}{}^{\alpha\beta}$.  
In particular, the first-order term is given by 
\be
\oone{\epsilon}^{}_{\mu}{}^{\sigma\alpha\beta} = -h^{\sigma}_{\lambda}\,
\bar\epsilon^{}_{\mu}{}^{\lambda\alpha\beta} + 2\,\bar\epsilon^{}_{\mu}{}^{\sigma}{}^{}_{\lambda}
{}^{[\alpha}h^{\beta]\lambda} 
+ \frac{1}{2}h \,\bar{\epsilon}^{}_{\mu}{}^{\sigma\alpha\beta}\,. \label{epsilon1}
\ee
The expansion of the second piece of the Cotton tensor, $C^{2}_{\mu\nu}$, 
is given by:
\begin{widetext}
\be
\oone C^{2}_{\mu\nu} = \frac{1}{2}\,\vartheta^{}_{|\sigma\tau}\left[
\bar{\epsilon}^{}_{(\mu}{}^{\sigma\alpha\beta}
\,\oone R^{\tau}{}^{}_{\nu)\alpha\beta} + \oone 
{\epsilon}^{}_{(\mu}{}^{\sigma\alpha\beta} 
\bar{R}^{\tau}{}^{}_{\nu)\alpha\beta}\right] - \frac{1}{2} 
\vartheta^{}_{|\rho} \oone S^\rho_{\sigma\tau} {\epsilon}^{}_{(\mu}{}^{\sigma\alpha\beta} 
\bar{R}^{\tau}{}^{}_{\nu)\alpha\beta}\,, \label{1stC2}
\ee
\ba
\otwo C^{2}_{\mu\nu} & = & \frac{1}{2}\,\vartheta^{}_{|\sigma\tau}\left[
\bar{\epsilon}^{}_{(\mu}{}^{\sigma\alpha\beta}
\,\otwo R^{\tau}{}^{}_{\nu)\alpha\beta} + \oone 
{\epsilon}^{}_{(\mu}{}^{\sigma\alpha\beta} 
\oone R^{\tau}{}^{}_{\nu)\alpha\beta}+ \otwo 
{\epsilon}^{}_{(\mu}{}^{\sigma\alpha\beta} 
\bar{R}^{\tau}{}^{}_{\nu)\alpha\beta}\right] \nn \\
& - & \frac{1}{2} 
\vartheta^{}_{|\rho} \oone S^\rho_{\sigma\tau}\left[
\bar{\epsilon}^{}_{(\mu}{}^{\sigma\alpha\beta}
\,\oone R^{\tau}{}^{}_{\nu)\alpha\beta} + \oone 
{\epsilon}^{}_{(\mu}{}^{\sigma\alpha\beta} 
\bar{R}^{\tau}{}^{}_{\nu)\alpha\beta}\right] 
- \frac{1}{2} 
\vartheta^{}_{|\rho} \oone S^\rho_{\sigma\tau}\,
{\bar\epsilon}^{}_{(\mu}{}^{\sigma\alpha\beta} 
\bar{R}^{\tau}{}^{}_{\nu)\alpha\beta}\,,
\ea
\end{widetext}
%

%---------------------------------------------------------------------------------------------------------------
\subsection{Propagation Equations in the SWA} \label{linearswa}
The propagation of oscillations in DCSMG using the SWA approximation is described by
the first-order equations given in Eq.~\eqref{swagwprop}.   Here, we analyze the 
different terms that appear in this equation adopting the TT gauge defined by 
Eqs.~\eqref{lorenzgauge} and~\eqref{tracelesscondition}.  We also make
use of the SWA to simplify the form of some of the terms.  Then, taking into
account that
\be
\oone S^{\alpha}_{\alpha\mu} = 0\,, \qquad
\bar{\met}^{\mu\nu}\,\oone S^\alpha_{\mu\nu} =0\,,
\ee
the first term in~\eqref{swagwprop} can be written as
\be
\oone R^{}_{\mu\nu} = \oone S^{\alpha}_{\mu\nu|\alpha} = -\frac{1}{2}h^{}_{\mu\nu|\rho}{}^{\rho}
+ {\cal O}\left(\frac{h^{}_{\GW}}{{\cal R}^2} \right)\,. \label{divS1}
\ee
From Eq.~\eqref{Ctensor} we have 
\ba
C_{\mu\nu}^{1}[\tilde\vartheta,\bar\met] &=& \tilde\vartheta^{}_{|\sigma}\,
\bar\epsilon^{\,\sigma\alpha\beta}{}^{}_{(\mu}\bar{R}^{}_{\nu)\alpha|\beta}
 = {\cal O}\left(\frac{\tilde\vartheta^{}_{\GW}}{\lambda^{}_{\GW}{\cal R}^{3}}\right)\,, \\
C_{\mu\nu}^{2}[\tilde\vartheta,\bar\met] &=& \tilde\vartheta^{}_{|\sigma\delta} 
{\,^\ast\!}\bar{R}^{\delta}{}^{}_{(\mu\nu)}{}^{\sigma}
 = {\cal O}\left(\frac{\tilde\vartheta^{}_{\GW}}{\lambda^{2}_{\GW}{\cal R}^{2}}\right)\,.
\ea
where $\tilde\vartheta^{}_{\GW}$ denotes the amplitude of the oscillations in the CS scalar
field.  From Eqs.~\eqref{1stC1} and~\eqref{1stC2} we can write:
\be
\oone C^{1}_{\mu\nu}[\bar\vartheta,h] = \left[ \bar\vartheta^{}_{|\sigma}
\bar{\epsilon}^{}_{(\mu}{}^{\sigma\alpha\beta}\,\oone S^{\rho}_{\nu)\beta|\rho}\right]_{|\alpha} 
+ {\cal O}\left(\frac{h^{}_{\GW}\,\bar\vartheta^{}_{b}}{\lgw{\cal R}^3} \right) \,,
\ee
\be
\oone C^{2}_{\mu\nu}[\bar\vartheta,h] = \bar\vartheta^{}_{|\sigma\tau} 
\bar{\epsilon}^{}_{(\mu}{}^{\sigma\alpha\beta}\,\oone S^{\tau}_{\nu)\beta|\alpha} 
+ {\cal O}\left(\frac{h^{}_{\GW}\,\bar\vartheta^{}_{b}}{\lgw{\cal R}^3} \right) \,,
\ee
where $\bar\vartheta^{}_{b}$ denotes the magnitude of the background value of the CS scalar field.
The last term in Eq.~\eqref{swagwprop} that we have to look at is the variation in the 
stress-energy tensor of the CS scalar field, 
$\delta T^{(\vartheta)}_{\mu\nu}[\bar\vartheta,\tilde\vartheta]$, which is given by
\be
\delta T^{(\vartheta)}_{\mu\nu}[\bar\vartheta,\tilde\vartheta] = 2\beta\left[ 
\bar\vartheta^{}_{|(\mu} \tilde\vartheta^{}_{|\nu)} - \frac{1}{2}\bar\met^{}_{\mu\nu}\,
\bar\vartheta^{|\sigma} \tilde\vartheta^{}_{|\sigma}  \right]\,.
\ee
Then, the first-order propagation equations [Eqs.~\eqref{swagwprop}], assuming for simplicity that
the waves travel in the absence of matter fields, can be written as
\ba
& &\left\{ \oone S^\alpha_{\mu\nu} + \frac{\alpha}{\kappa}\left[\bar\vartheta^{}_{|\sigma}
\bar{\epsilon}^{}_{(\mu}{}^{\sigma\alpha\beta}\,\oone S^{\rho}_{\nu)\beta} \right]_{|\rho}
\right\}_{|\alpha}  = \frac{\beta}{\kappa}\, \bar\vartheta^{}_{|(\mu} \tilde\vartheta^{}_{|\nu)} 
\nonumber \\
& & +\;{\cal O}\left(\frac{h^{}_{\GW}\,\bar\vartheta^{}_{b}}{\lgw{\cal R}^3}\,,
\frac{\tilde\vartheta^{}_{\GW}}{\lambda^{}_{\GW}{\cal R}^{3}}\,,
\frac{\tilde\vartheta^{}_{\GW}}{\lambda^{2}_{\GW}{\cal R}^{2}}\right) \,.
\label{firstorderequations}
\ea
As it is clear, this propagation equation for 
$h^{}_{\mu\nu}$ is coupled to the propagation equation for $\tilde\vartheta$ 
[Eq.~\eqref{swacsprop}].  Notice that the right-hand side of the equations can be written 
as a total derivative, both in 
GR and in DCSMG. Also note that the second term 
in the right-hand side is to leading-order of 
${\cal O}(h^{}_{\GW}\,\bar\vartheta^{}_{b}/(\lambda^{3}_{\GW}{\cal R}))$.

%%%%%%%%%%%%%%%%%%%%%%%%%%%%%%%%%%%%%%%%%%%%%%%%%%%%%%%%%%%%%%%%%%%%%%%%%%%%%%

\end{document}